\documentstyle[emulateapj,pstricks]{article}
\def\clone{\mbox{$Cl1_{0.0}$}}
\def\cltwo{\mbox{$Cl2_{0.0}$}}
\def\clthree{\mbox{$Cl1_{0.1}$}}
\def\clfour{\mbox{$Cl1_{0.5}$}}
\def\clfive{\mbox{$Cl1_{3.0}$}}
\def\clsix{\mbox{$Cl2_{3.0}$}}
\def\clseven{\mbox{$Cl1_{0.5/V}$}}
\def\cleight{\mbox{$Cl1_{1.0/V}$}}
\def\cmgr{\mbox{cm$^2$gr$^{-1}$}}
\def\densunit{\mbox{h$^2$M$_{\odot}$Mpc$^{-3}$}}
\def\densunitF{\mbox{M$_{\odot}$pc$^{-3}$}}

\def\Gone{\mbox{$G1_{0.0}$}}
\def\Gtwo{\mbox{$G2_{0.0}$}}
\def\Gthree{\mbox{$G3_{0.0}$}}
\def\Gfour{\mbox{$G1_{0.1}$}}
\def\Gfive{\mbox{$G1_{0.5}$}}
\def\Gsix{\mbox{$G2_{0.5}$}}
\def\Gseven{\mbox{$G4_{0.5}$}}
\def\Geight{\mbox{$G5_{0.5}$}}
\def\Gnine{\mbox{$G1_{3.0}$}}
\def\Gten{\mbox{$G2_{3.0}$}}
\def\Geleven{\mbox{$G3_{3.0}$}}
\def\Gtwelve{\mbox{$G1_{0.5/V}$}}
\def\Gthirteen{\mbox{$G2_{0.5/V}$}}
\def\Gfourteen{\mbox{$G1_{1.0/V}$}}
\def\hubble{\mbox{km s$^{-1}$ Mpc$^{-1}$}}

\def\kms{\mbox{kms$^{-1}$}}
\def\kpch{\mbox{h$^{-1}$kpc}}

\def\mpch{\mbox{h$^{-1}$Mpc}}
\def\msun{\mbox{M$_\odot$}}     
\def\msunh{\mbox{h$^{-1}$M$_\odot$}}

\def\mvir{\mbox{M$_{vir}$}}
\def\ome{\mbox{$\Omega_m$}}

\def\omel{\mbox{$\Omega_\Lambda$}}
\def\rcsl{\mbox{r$_{c,-1}$}}
\def\rhoc{\mbox{$\rho_c$}}
\def\rhosl{\mbox{$\rho_{c,-1}$}}
\def\rvir{\mbox{R$_{vir}$}}
\def\sig{\mbox{$\sigma_{DM}$}}
\def\sige{\mbox{$\sigma_8$}}
\def\siga{\mbox{$\hat{\sigma}$}}
\def\sign{\mbox{$\sigma_0$}}
\def\tdy{\mbox{t$_{dyn}$}}
\def\vcien{\mbox{v$_{100}$}}
\def\vmax{\mbox{v$_{\rm max}$}}
\def\vrel{\mbox{v$_{12}$}}
\def\ncol{\mbox{N$_{\rm col}$}}
\def\mathnew{\mathsurround=0pt}
\def\ref{\par\noindent\hangindent=2pc \hangafter=1 }
\def\simov#1#2{\lower .5pt\vbox{\baselineskip0pt
    \lineskip-.5pt\ialign{$\mathnew#1\hfil##\hfil$\crcr#2\crcr\sim\crcr}}}  

\def\simless{\mathrel{\mathpalette\simov <}}
\def\'#1{\ifx#1i{\accent"13\i}\else{\accent"13#1}\fi}
\def\eg{e.g.,}

\def\ie{i.e.}
\begin{document}
\slugcomment{{\em accepted by the Astrophysical Journal}}

\lefthead{HALOS IN A SELF-INTERACTING DARK MATTER COSMOLOGY}
\righthead{COLIN ET AL.}

\title{Structure and Subhalo Population of Halos in a Self-Interacting 
Dark Matter Cosmology}

\author{Pedro Col\'in and Vladimir Avila-Reese,}
\affil{Instituto de Astronom\'ia, Universidad Nacional Aut\'onoma
de M\'exico, A.P. 70-264, 04510, M\'exico, D.F., M\'exico}

\author{Octavio Valenzuela}
\affil{Astronomy Department, New Mexico State University, Box 30001, 
Department 4500, Las Cruces, NM 88003-0001}

\author{and} 

\author{Claudio Firmani}
\affil{Osservatorio Astronomico di Brera, via E. Bianchi 46, 23807 Merate
 (LC), Italy}


\begin{abstract}
A series of high-resolution numerical simulations were performed
to study the structure and substructure of Milky Way (MW)- and cluster-sized 
halos in a $\Lambda-$Cold Dark Matter (CDM) cosmology with self-interacting 
(SI) dark particles. The cross section per unit of particle mass has the form 
$\sig = \sign (1/ \vcien)^\alpha$, where \sign\ is a constant in units 
of \cmgr\ and $\vcien$ is the relative velocity in units of 100 \kms.  
Different values for $\sign$ with $\alpha= 0$ or 1 were used. For small 
values of \sig = const. ($\lesssim 0.5$, $\alpha=0$), the core density 
of the halos at $z=0$ is typically higher at a given mass for lower values 
of \sign\ or, at a given \sign, for lower masses. For values of 
$\sign$ as high as 3.0, both cluster- and MW-sized halos may undergo the 
gravothermal catastrophe before $z=0$. The core expansion occurs in a stable 
regime because the heat capacity, $C$, is positive in the center. After the 
maximum expansion, the isothermal core is hotter than the periphery and 
$C<0$. Then, the gravothermal catastrophe triggers. The instability onset 
can be delayed by both the dynamical heating of the halo by major mergers 
and the interaction of cool particles with the hot environment of a host 
halo. When $\alpha = 1$, the core density of cluster- and MW-sized halos 
is similar. Using $\sig = 0.5-1.0\ (1/\vcien)$, our predictions agree with 
the central densities and the core scaling laws of halos both inferred 
from the observations of dwarf and low surface brightness galaxies and 
clusters of 
galaxies. Regarding the cumulative $\vmax-$function of subhalos within 
MW-sized halos, when ($\sign,\alpha$) = (0.1,0.0),  (0.5,0.0) or (0.5,1.0) 
it agrees roughly with observations (luminous satellites) for
$\vmax \gtrsim 30 \kms$, while at $\vmax = 20 \kms$ the functions are 
already a factor 5-8 higher, similar to the CDM predictions. For 
($\sign,\alpha$) = (1.0,1.0), this function lies above the corresponding 
CDM function. The structure and number of subhalos are affected by the 
scattering properties of the host halo rather than by those of the subhalos. 
The halos with SI have more specific angular momentum at a given mass shell 
and are rounder than their CDM counterparts. However, the angular momentum 
excess w.r.t. CDM is small. We conclude that the introduction of SI particles 
with $\sig \propto 1/\vcien$ {\it may remedy the cuspy core problem} of 
the CDM cosmogony, at the same time keeping a {\it subhalo population 
similar to that of the CDM halos}. 

\end{abstract}
 
%=====================

\section{Introduction}

%=====================

The current paradigm for cosmic structure formation is 
based on the cold dark matter (CDM) cosmological models, 
where CDM particles are assumed to be collisionless. This 
is the simplest assumption about the nature of these particles, 
not yet detected. The predictions of the most popular encarnation 
of the CDM variants, the spatially flat with a non-vanishing 
cosmological constant model, are in excellent agreement with a 
large body of observational data at large scales. With the advent 
of high-resolution numerical simulations and new observational 
techniques, a comparison between models and observations at small 
scales has become possible. Apparent conflicts have 
emerged as a result of this comparison: (i) the predicted halo inner 
density profiles are cuspy, in disagreement with the shallow 
cores favored by observations of dwarf and low surface
brightness galaxies (LSB), and (ii) the predicted number of subhalos 
within Milky Way-sized (MW-sized) halos overwhelm the 
observed abundance of satellite galaxies in the Local Group.

Recently, a plethora of alternative theories, which modifiy the 
predictions of the CDM model at small scales but retain its successes 
at large scales, have been proposed (\eg\ Dav\'e et al. 2001 and 
references there included). In one of these scenarios, CDM particles are 
assumed to be self-interacting (SI) in such a way that the heat flux to
the core smooth out the density cusp, and simultaneously reduces the 
amount of substructure by evaporating orbiting subhalos (\cite{SS00}).
Several authors have explored this model numerically and 
analytically and concluded that the relevant regime for structure 
formation would have to be the optically thin one (\cite{Moore00}; 
Yoshida et al. 2000a; Firmani et al. 2001b; \cite{KW}; Dav\'e 
et al. 2001; \cite{HO01}; \cite{BSI02}). By means of cosmological N-body 
simulations, using a {\it constant} cross section per unit mass, \sig, 
Dav\'e et al. (2001) found that halos at galaxy scales have long-lived 
shallow cores whose sizes agree with observational inferences when 
$\sig \approx 5\ \cmgr$. At the cluster scales, Yoshida et al. (2000b) 
also found that the SIDM halos present long-lived shallow cores. According 
to them, \sig\ should be smaller than $\approx 5\ \cmgr$ in order for 
their results to be in agreement with observations of cluster of galaxies. 

On the other hand, N-body simulations of relaxed, isolated, SI dark matter 
(SIDM)  halos with initial cuspy density profiles and with a relatively 
high \sig\ value show a core evolution which is remarkably rapid. After
the maximum soft core radius is reached, in time scales of the order of
the halo dynamical time, the core collapse is triggered almost immediately  
(\cite{KW}, hereafter KW; Burkert 2000; see also \cite{Quinlan96}). 
\cite{Dave2001} argue that the fast core collapse seen in these simulations 
may be attributed to the absence, in the isolated case, of infalling 
dynamically-hot material. Dav\'e et al. also find that the central density 
of simulated cosmological halos (at 1 \kpch\ from the center) remains 
approximately constant with halo mass, a result actually not expected for a 
constant \sig\ (Yoshida et al. 2000b; Firmani et al. 2001a,b). In fact, 
Firmani et al. (2001b) find that 
\rhoc\ is nearly independent of halo mass only when \sig\ is {\it inversely 
proportional to the relative velocity of colliding particles}. They 
used a numerical code based on the collisional Boltzman equation and took 
into account the cosmological mass aggregation process during the formation 
of their spherical symmetric halos. They obtain a central density of 
$\rhoc \approx 0.02\densunitF$  (a value inferred from observations, 
$h=0.65$ was assumed, Firmani et al. 2000a) for $\sig \approx (0.6 /\vcien) \ \cmgr$, 
where \vcien\ is the relative velocity \vrel\ in units of 100 \kms. 

As suggested by all these studies, important questions regarding the SIDM 
cosmology still remain without a satisfactory answer or without an answer at 
all. Do the SIDM cosmological halos undergo a core collapse on scales times 
shorter than a Hubble time? How does the cosmological mass aggregation 
history (MAH) affect the core evolution? Does a constant cross section
produce halos with a nearly constant central density regardless of their mass?
Which are the predictions for SIDM models with a cross section inversely 
proportional to \vrel?
How is the subhalo population at cluster and galaxy scales
in a SIDM cosmology? In this paper, we will address these questions using
high-resolution cosmological simulations of cluster- and galaxy-sized 
halos with $\sig\ \propto\ $const and $\propto1/\vrel$. 
  
Our study is aimed to explore whether the SIDM model predictions are in 
better agreement with observations than the CDM ones. As already disscused 
above, the key parameter for SIDM is the interaction cross section, 
\begin{equation}
\sigma_{DM} = \sigma_0 \left( \frac{1}{\vrel} \right)^\alpha. 
\end{equation}
A number of authors have attempted to constrain the range of values for 
the pair of parameters ($\sign,\alpha$) (Firmani et al. 2000, 2001a; 
Miralda-Escud\'e 2000; \cite{Wandelt00}; \cite{Meneghetti01}; 
\cite{GO2001}; \cite{HO01}). According to analytical and semi-analytical 
arguments of Hennawi \& Ostriker (2001; HO hereafter), \sig\ should be 
large enough to produce a shallow core in agreement with observations, 
but small enough so the core halo would not collapse in a Hubble time 
or subhalo evaporation within cluster-sized halos would not alter 
dramatically the elliptical Fundamental Plane relations. This leaves 
the SIDM cosmology with a limited range of ($\sign,\alpha$) values. 
HO also introduce an extra strong constrain for SIDM, the mass of a 
central black hole in the MW and the lack of it in M33. According to 
them, accretion of SIDM onto seed black holes produces holes that can 
be much more massive than those observed if \sig\ is roughly as large 
as needed to form the soft cores inferred from observations. In \S 5.2 
we will discuss this constrain in light of our results.

In \S 2 we describe the method and strategy to simulate SIDM halos,
and present the simulation of a monolithical halo (\S 2.1) as well as 
the cosmological simulations carried out in this paper (\S 2.2). In 
\S 3.1 we analyze the density profiles and central densities measured 
in our simulated cluster- and galaxy-sized halos with different values of 
($\sign,\alpha$) and different merging histories. In \S 3.2 we 
discuss the concentration parameters of halos and subhalos, while in
\S 3.3 and 3.4 we present results regarding the ellipticity and 
angular momentum distribution of the SIDM halos. Section 4 is devoted to 
the study of the subhalo population in the MW- and cluster-sized SIDM 
halos. In \S 5 we discuss our results and the viability of the SIDM models,
and in \S 6 we summarize this paper and present our conclusions.

%=====================

\section{Simulations}

%=====================

The Adaptive Refinement Tree (ART) N-body code (\cite{KKK97})
has been used to run the N-body simulations. 
The ART code achieves high spatial resolution 
by refining the base uniform grid in all high-density regions 
with an automated refinement algorithm. The elastic collisions,
characterized by a scattering cross section $\sig$, are implemented
in ART in the following manner. In a medium with density $\rho$, the 
predicted distance $d$ that a particle travels without collisioning 
is given by (Gibbs 1994)
\begin{equation} 
d = - \lambda \ln(1 - P),
\end{equation} 
where $\lambda = 1 / \rho \sig$ is the mean free path and $P$ is 
a random number distributed uniformly between 0 and 1. A pair of 
particles collide if this distance $d$ results lower than the distance
$\vrel \Delta t$, where \vrel\ is the relative velocity between the
particle and {\it one of} its nearest neighbors, and $\Delta t$ is the 
time step. In a inhomogenous medium, we substitute $\rho$ by the local
density at the particle's position, $\rho_i$. We exploit the mesh
hierarchy structure of ART to evaluate both the local density and the 
relative velocity. On each level of the structure, starting from the 
finest level, ART assigns the density in each {\it cell} using
the standard cloud-in-cell technique (\cite{HE1981}). The density at 
the particle's position is found by interpolation\footnote{In ART, only 
those cells that are not refined are allowed to own particles} using 
cloud-in-cell. The search for a partner to collide with may involve many 
neighbor cells and more than one level. We find that the search for 
partners of particles that lie in low-density regions fails often;
this, however, does not affect our results because collisions in these 
regions are very rare. It happens many times that the partner 
is found in the same cell where the particle is, and thus the search is
limited to only one cell. The new velocities after collision are computed 
by imposing energy and momentum conservation and by orienting the 
after-collision relative velocity randomly.
{\pspicture(-0.6,-1.0)(12.0,17.0)
\rput[tl]{0}(-0.5,16.5){\epsfxsize=8.5cm
\epsffile{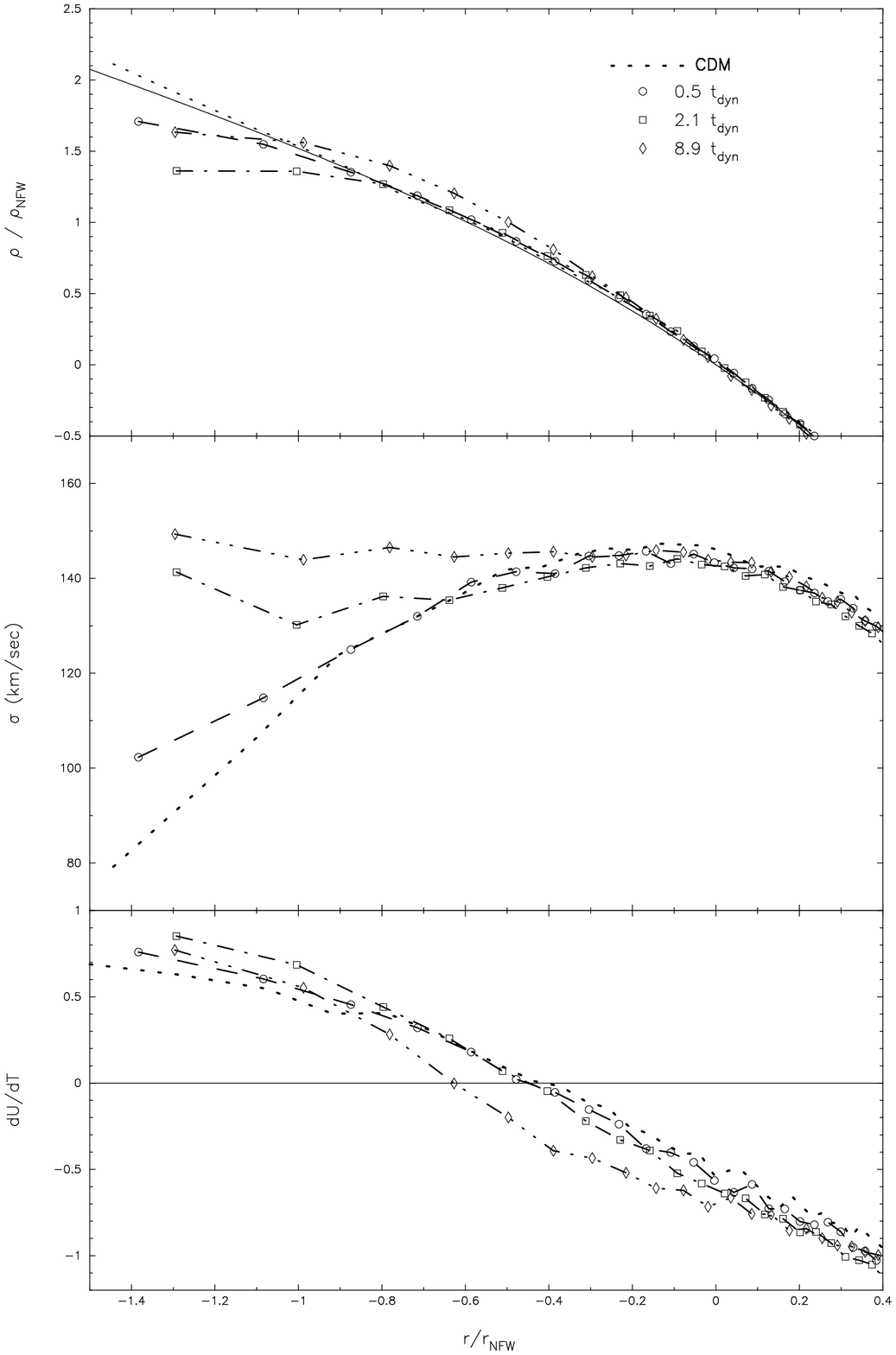}}
\rput[tl]{0}(-0.7,2.4){
\begin{minipage}{8.8cm}
  \small\parindent=3.5mm {\sc Fig.}~1.---
Evolution of the density, 3D velocity dispersion, and heat 
capacity profiles (top, central and bottom panels, respectively) for the NFW 
isolated SIDM halo of mass $2 \times 10^{11}\ \msunh$ and initial concentration 
$c_{\rm NFW} = 14$. Tha maximum expansion of the core occurs at about 
$2 \tdy$. With a solid line  
is represented the analytical NFW function from which a relaxed and stable dark
matter halo is reproduced (dotted line). We see that at 9 \tdy\ the central
density has increased by almost a factor 2.
\end{minipage}}
\endpspicture}

%=====================

\subsection{Self-Interacting simulations of a monolithic halo}

%=====================

In order to test our collisional algorithm on ART code we ran a simulation
of a monolithic (isolated) halo of $3 \times 10^5$ particles with an initial 
NFW density profile, and SI turned on. The virial mass of the halo is 
$\mvir = 2 \times 10^{11}\ \msunh$, with a NFW concentration of
$c_{\rm NFW} = 14$. If we assume the Hernquist model to have the same 
density as the NFW one in the central $r^{-1}$ region, we can relate 
the scale radius $r_H$ of the Hernquist density profile with $c_{\rm NFW}$. 
Given $r_H$ and \sig\ we can estimate the dimensionless cross section \siga\ 
and the core relaxation time $t_{rc}$, both as defined by Kochanek \& White 
(2000, hereafter KW) who also simulated SIDM monolithic 
halos but with a Hernquist profile. The 
simulation was run when the SI algorithm was still
on test and used the particle velocity for the computation of the collisional
probability instead of \vrel; this produces a smaller scattering rate
than when one uses \vrel\ (KW). The model has $\sig = 9.0\ \cmgr$. 

The evolution of the density, 3D velocity dispersion, and heat capacity 
profiles are shown in Figure 1; time 
is measured in units of the dynamical time, $t_{\rm dyn}$, as defined 
in KW. The evolution proceeds in much the same way as in previous 
similar SI simulations (e.g., Burkert 2000; KW). If the initial equilibrium 
configuration of a halo has a NFW density profile, the velocity 
dispersion (temperature) profile peaks out of the center, close to the 
scale radius r$_{\rm NFW}$ (dotted line in Fig. 1); i.e., there is an i
nversion of the temperature profile. We have measured the potential, 
kinetic, and total energy, $W$, $T$, and $U$, respectively, of our 
monolithic halo as a function of radius and found, for the initial 
configuration, that both the total energy (see also Fig. 2 of \cite{LM2001}) 
and the heat capacity, $C$, of the innermost region are positive (see Fig. 1).

Therefore, when the SI is turned on, heat starts flowing into the core 
from the immediate surroundings and the core expands. An increase in the
core's total energy produces an increase in the core's temperature and 
in the ratio $2T/|W|$; in fact, the maximum expansion of the core occurs
approximately when  $2T/|W|$ also reaches its maximum. At this point the
core is isothermal and the temperature inversion is gone (Fig. 1; see 
also panel (a) in Fig. 2). Now, the isothermal core is hotter than the 
surroundings and heat starts flowing outwards (note that this process 
happens in the outer central region where $C$ is already negative). 
The central region, where both $U$ and $C$ are positive, reduces in size
(Fig. 1). 
The core looses heat and contracts; the system is unstable and the 
gravothermal catastrophe phase triggers (see Binney \& Tremaine 1987, 
\S 8.2): the core becomes hotter as it looses energy outwards, and the
increased temperature difference leads to an even faster energy loss with 
the consequent core shrinking (panel (b) in Fig. 2). 

An interesting question is whether the inverse instability, the gravothermal 
expansion, may arise sometime in the evolution of a halo. This
bring us to the question: why at the begining of the SI evolution of
the isolated halo, when there is an inversion of the temperature profile, a
run-away core expansion does not occur? The reason is that the central region
has positive total energy and heat capacity. Let us now analize the halo 
after the maximum core expansion and let us assume that for some reason 
(for example, kinetic energy injection by mergers), the outer region 
of a halo becomes hotter than the isothermal core (panel (c) in Fig. 2).  
The system in this case does not obey the conditions for the gravothermal 
expansion instability because, again, $U$ and $C$ in the central region are 
positive (due to the strong halo heating, $C$ may become positive even at the 
periphery). A flow of heat into the core makes it expand and brings most 
parts of the system to an isothermal state (panel (c)), which might be 
unstable to core collapse because the isothermal sphere becomes a local 
entropy minimum rather than a local entropy maximum. Thus, the core 
collapse phase for cosmological SIDM halos can be delayed due to an external 
energy injection, {\it but not reversed.} We will return to this 
disscusion in \S 3 where we analyze cosmological SIDM halos.
{\pspicture(-0.6,-0.5)(12.0,16.0)
\rput[tl]{0}(-0.5,15.5){\epsfxsize=8.5cm
\epsffile{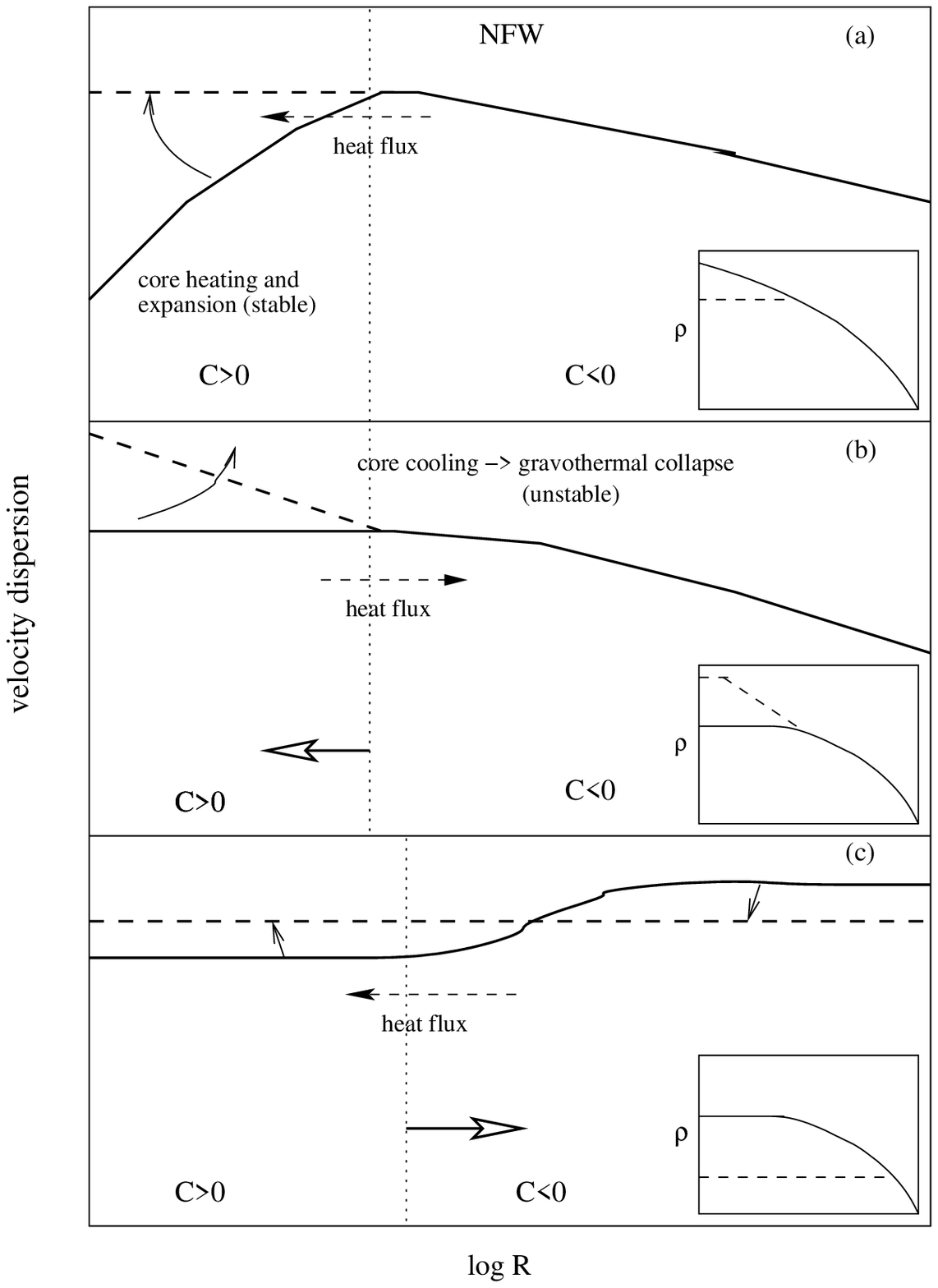}}
\rput[tl]{0}(-0.7,3.0){
\begin{minipage}{8.8cm}
  \small\parindent=3.5mm {\sc Fig.}~2.---
Schematical behavior of the 3D velocity dispersion 
(temperature)
profile in a SIDM halo, and of the logarithmic density profile (encapsulated
boxes). Panels (a), (b), and (c) show the core expansion phase (stable),
the gravothermal catastrophe phase (unstable), and the core collapse delay
(and core expansion) when the halo is externally heated, respectively. See the 
text for more explanations. $C$ is the heat capacity.
\end{minipage}}
\endpspicture}

The notion of gravothermal catastrophe has been introduced for a non-singular 
isothermal model (Lynden-Bell \& Wood 1968). Although in the cosmological case
the halos are not described by isothermal spheres, the core expansion phase 
leads a significant part of the SIDM halo to a state close to that of a 
non-singular sphere. This is for example the case of the halo at maximum 
expansion in Fig. 1 (squares) where r$_c\approx (9\sigma^2/4\pi G\rho_c)^{1/2}$ and 
r$_c$ is roughly equal to the radius where $C$ is still positive. 
Interestingly, the state of this halo is close to the non-equilibrium state 
of minimum entropy described by Lynden-Bell \& Wood (1968). At the radius 
where the velocity dispersion (temperature) is still roughly constant 
(the radius of the isothermal sphere), the density has decreased from the 
center by approximately a factor of 30 and $E$ and $C$ are already negative. 
Systems like this are susceptible to the gravothermal catastrophe 
and this is what we observe it happens in our simulations. 

For our monolithic NFW halo, the maximum constant density core radius 
is reached approximately in $2.6 \times 10^9$ yr. This time scale agrees 
with the estimated core relaxation time, $t_{\rm rc}$, by KW for 
$\siga \sim 0.8$ in their dimensionless units or, equivalently, 
$\sig \sim 5.0\ \cmgr$. As mentioned above, had we used \vrel\ 
instead of the particle velocity, a higher scattering rate would have 
been expected. The factor 9/5, however, is lower than the factor 3 that 
KW claim they measure. The difference is probably due to the way in which 
we compute local density. If we estimate the core radius $r_c$ by the 
point where the density drops to 1/4 the central density, then 
$r_c \sim 0.4 r_{\rm NFW}$. Using the same definition for $r_c$, 
\cite{Burkert2000} finds $r_c \sim 0.6 r_p$ , where $r_p$ is the radius 
where the velocity dispersion peaks for a Hernquist density profile. 
Because the radius $r_p$ is close to $r_{\rm NFW}$ for a NFW density 
profile, we would have expected, for pure analogy with Burkert's results, 
that $r_c \sim 0.6 r_{\rm NFW}$. We see from Figure 1, panel (b), that the 
peak is broad and is slightly lower than $r_{\rm NFW}$. It is not thus 
difficult to see why the core size is not as large as in the Hernquist's 
case. We let the simulation run for another $8.5 \times 10^9$ yr and compared 
the rate at which the core collapse is produced with that predicted by KW 
(see their Fig. 2). In the KW halo,  for which $t_{\rm rc} \sim  2.6 
\times 10^9$ yr and $\siga \sim 0.8$, the core collapse develops at a faster 
rate than the
one we observe in our halo. Our core density evolution is more akin to a KW 
\siga\ simulation with a value between 1.0 and 3.0. It is not clear what this 
difference is due to, but this difference could also be explained by the fact 
that we are using a NFW density profile instead of a Hernquist one. In any 
case, we find a good agreement between our results and Burkert's and KW 
results. 

%=====================

\subsection{Cosmological simulations of SIDM halos}

%=====================

The set of simulations performed to analyze the structure of halos in 
a SIDM cosmology uses the multiple-mass variant of ART (Klypin et al. 2001)
in order to increase the mass and spatial resolution in 
few selected halos. All models assume a $\Lambda$CDM power spectrum
with total matter density and cosmological constant, in units of the 
critical density, of $\ome = 0.3$ and $\omel = 0.7$, a Hubble constant 
of $ h = 0.7$, in units of $100\ \hubble$, and a $\sige = 1.0$. Here 
\sige\ represents the rms of mass fluctuations estimated with the t
op-hat window of radius $8 \mpch$.

We first perform two low-mass resolution simulations, one in a 100 \mpch\ 
box and the other in a 12.5 \mpch\ box. Both simulations were run
with $64^3$ particles in a grid initially consisting of $256^3$ cubic 
cells. Two cluster-sized halos from the 100 \mpch\ box simulation
and five MW-sized halos from the 12.5\mpch\ box simulation
were selected for being resimulated with higher resolution. 
To increase resolution in the chosen halos, particles within 2.5 virial
radii from the halo centers were traced back to the initial epoch and 
new initial conditions were then generated by using the multiple-mass
scheme described in \cite{KKBP00}. Three different mass levels with
1, 8, and 64 times $m_p$ were used, where $m_p$ is the mass per particle
at the finest mass level (see Table 1). Only particles at this level
were allowed to collide. Final halos have a few $10^5$ particles each 
and contamination within their virial radii by heavy particles was 
kept below 1\%. We notice that initial conditions do not change when
\sig\ is varied.
\begin{planotable}{rrccccrcrc}
\tablewidth{0pt}
\tablehead{\colhead{$L_{BOX}$} & \colhead{$m_p$} & \colhead{Resolution} & 
\colhead{$\sign$} & \colhead{$\alpha$} & \colhead{$M_{vir}$} & \colhead{$v_{max}$} & 
\colhead{$\lambda'$} & \colhead{\ncol} &
\colhead{Halo name tag} \\
(\mpch) &  (\msunh) & (\kpch) & (\cmgr) &  & (\msunh) & (\kms) & ($10^{-2}$) &  & }
\startdata
100.0 & $5.0 \times 10^9$ & 3.0 & 0.0 & 0.0 & $4.5 \times 10^{14}$ & 1280 & 4.43 & 0.00 & \clone\ \\
100.0 & $5.0 \times 10^9$ & 3.0 & 0.0 & 0.0 & $5.0 \times 10^{14}$ & 1230 & 1.80 & 0.00 & \cltwo\ \\
100.0 & $5.0 \times 10^9$ & 6.1 & 0.1 & 0.0 & $4.4 \times 10^{14}$ & 1285 & 4.57 & 0.60 & \clthree\ \\
100.0 & $5.0 \times 10^9$ & 6.1 & 0.5 & 0.0 & $4.5 \times 10^{14}$ & 1299 & 4.88 & 2.37 & \clfour\ \\
100.0 & $5.0 \times 10^9$ & 6.1 & 3.0 & 0.0 & $4.4 \times 10^{14}$ & 1336 & 4.62 &12.54 & \clfive\ \\
100.0 & $5.0 \times 10^9$ & 3.0 & 3.0 & 0.0 & $3.6 \times 10^{14}$ & 1548 & 2.17 &30.03 & \clsix\ \\
100.0 & $5.0 \times 10^9$ & 6.1 & 0.5 & 1.0 & $4.4 \times 10^{14}$ & 1276 & 4.47 & 0.35 & \clseven\ \\
100.0 & $5.0 \times 10^9$ & 6.1 & 1.0 & 1.0 & $4.4 \times 10^{14}$ & 1270 & 4.45 & 0.64 & \cleight\ \\
12.5 & $9.7 \times 10^6$ & 0.2 & 0.0 & 0.0 & $1.4 \times 10^{12}$ & 215 & 3.82 & 0.00 & \Gone\ \\
12.5 & $9.7 \times 10^6$ & 0.2 & 0.0 & 0.0 & $2.7 \times 10^{12}$ & 237 & 7.21 & 0.00 & \Gtwo\ \\
12.5 & $9.7 \times 10^6$ & 0.2 & 0.0 & 0.0 & $8.7 \times 10^{11}$ & 167 & 6.90 & 0.00 & \Gthree\ \\
12.5 & $9.7 \times 10^6$ & 0.4 & 0.1 & 0.0 & $1.3 \times 10^{12}$ & 214 & 4.18 & 0.33 & \Gfour\ \\
12.5 & $9.7 \times 10^6$ & 0.4 & 0.5 & 0.0 & $1.3 \times 10^{12}$ & 217 & 4.45 & 1.32 & \Gfive\ \\
12.5 & $9.7 \times 10^6$ & 0.4 & 0.5 & 0.0 & $2.6 \times 10^{12}$ & 240 & 7.45 & $\cdots$ & \Gsix\ \\
12.5 & $9.7 \times 10^6$ & 0.4 & 0.5 & 0.0 & $1.1 \times 10^{12}$ & 193 & 1.85 & $\cdots$ & \Gseven\ \\
12.5 & $9.7 \times 10^6$ & 0.4 & 0.5 & 0.0 & $1.1 \times 10^{12}$ & 178 & 5.53 & $\cdots$ & \Geight\ \\
12.5 & $9.7 \times 10^6$ & 0.4 & 3.0 & 0.0 & $1.2 \times 10^{12}$ & 245 & 4.50 & 9.41 & \Gnine\ \\
12.5 & $9.7 \times 10^6$ & 0.1 & 3.0 & 0.0 & $2.3 \times 10^{12}$ & 273 & 5.70 &15.03 & \Gten\ \\
12.5 & $9.7 \times 10^6$ & 0.4 & 3.0 & 0.0 & $8.3 \times 10^{11}$ & 161 & 6.02 & 4.02 & \Geleven\ \\
12.5 & $9.7 \times 10^6$ & 0.4 & 0.5 & 1.0 & $1.3 \times 10^{12}$ & 212 & 4.28 & 0.77 & \Gtwelve\ \\
12.5 & $9.7 \times 10^6$ & 0.4 & 0.5 & 1.0 & $2.7 \times 10^{12}$ & 239 & 7.53 & 0.95 & \Gthirteen\ \\
12.5 & $9.7 \times 10^6$ & 0.4 & 1.0 & 1.0 & $1.3 \times 10^{12}$ & 212 & 4.31 & 1.34 &\Gfourteen\ \\
\enddata
\end{planotable}

The bound density maxima (BDM) group finding algorithm (\cite{K99})
was used to identify halos and subhalos in all simulations. The BDM 
algorithm finds positions of local maxima in the density field 
smoothed at the scale of interest and applies physically motivated 
criteria to test whether a group of particles is a gravitationally bound 
halo. In the BDM algorithm, there is a parameter called the rejection velocity
limit $v_l$ which monitors whether a particle is bound or not. For standard 
CDM simulations, in which a virialized halo has a NFW like density profile,
we usually set $v_l = v_e$, where the escape velocity $v_e$ is analytically
evaluated in BDM assuming the halo follows a NFW density profile. We decided 
to turn off the rejection velocity switch because density profiles of SIDM 
halos are not supposed to be fitted by a NFW profile. For subhalos the switch 
is kept on for two reasons: first, a significant fraction of unbound 
particles is expected to populate these subhalos, and second, because these 
halos are composed of at most hundreds of particles, their profiles in most 
cases are well fitted by a NFW profile.
\subsection{Density Profiles}
\begin{figure*}[ht]
\pspicture(0.5,-1.5)(15.0,17.0)
\rput[tl]{0}(1.0,16.0){\epsfxsize=18cm
\epsffile{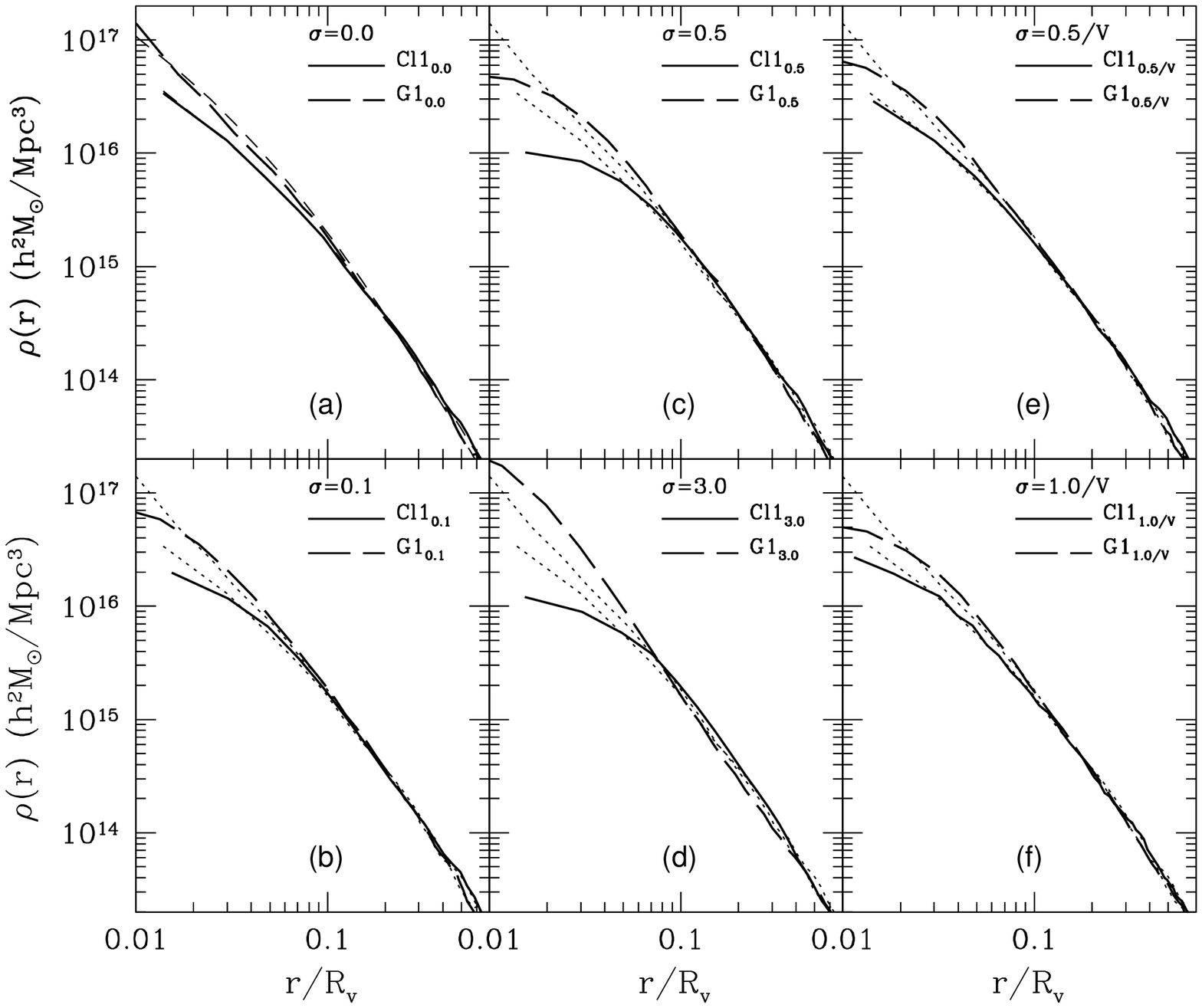}}
\rput[tl]{0}(0.5,0.0){
\begin{minipage}{18.4cm}
  \small\parindent=3.5mm {\sc Fig.}~3.---
Density profiles of the fiducial cluster- (solid lines) and 
MW-sized (dashed lines) halos simulated with different values of \sig\ 
(see top of each panel). V is the \vrel\ in unities of 100 \kms. Dotted 
curves are the profiles of 
the CDM case (panel a) plotted for comparison. Dashed curves in panel (a) are 
the best NFW fit to the profiles shown there. Radius is given in unities of 
the virial radius \rvir.
\end{minipage}}
\endpspicture
\end{figure*}

A summary of the parameters of the simulations of halos are 
presented in Table 1. The mass of the particle in the high-resolution 
region $m_p$ is shown in column (2) and is only a funtion of the size of 
the box (col. [1]) because the cosmological model and the number 
of mass levels are fixed. Column (3) shows the formal force resolution, 
measured by the size of a cell in the finest mesh. The collisionless CDM 
simulations have a higher resolution than the collisional ones because 
simulated halos in the former case reach a higher central density than in 
the latter one. Values of the parameters of the SI cross section, $\sigma_0$ 
and $\alpha$, are presented in columns (4) and (5). The mass \mvir\ within 
the virial radius, defined as the radius at which the average halo density 
is $\delta$ times the background density according to the spherical top-hat 
model, is shown in column (6). $\delta$ is a number that depends on epoch and 
cosmological parameters (\ome,\omel); for a flat $\Lambda$CDM model, 
$\delta \sim 337$ at $z = 0$, where $z$ is the redshift. The maximum circular 
velocity defined as
\begin{equation} 
        v_{\rm max} =\left( \frac{GM(<r)}{r} \right)^{1/2}_{\rm max},
\end{equation}
where $G$ is the gravitational constant and $M(<r)$ is the mass contained
within the radius $r$, is placed in column (7). The spin parameter $\lambda'$ 
as defined by Bullock et al. (2001) is shown in column (8) (see \S 3.4). In 
column (9) \ncol\ denotes the total number of collisions until $z = 0$
divided by the total number of particles within \rvir; as mentioned above, 
only particles at the finest mass-level of refinement are allowed to collide. 
This number could not be obtained for those halos (horizontal dots) that were 
simulated simultaneously in only one run because the collision counting refers 
to the whole region selected to be resimulated at high resolution. In column 
(10) we give the halo name tag for each one of the simulated halos. Labels 
$Cln$ denote cluster-sized halos ($n$ goes from 1 to 2), whereas labels 
$Gn$ denote MW-sized halos ($n$ goes from 1 to 5). The numbers $n=1,2$ for 
halos $Cl$, and $n=1,2,3$ for halos $G$ indicate different simulations
aimed to have halos with different mass assembly histories.  MW-sized halos
numbers 4 and 5 turned to appear in the same simulation as the halo $G2$,
and we have included them for completness. Both, the 
cluster- and MW-sized halos have been simulated for a variety of 
($\sign$,$\alpha$) values; the lower index in the name tag 
indicates the used combination. 
Note that only halos with $n=1$ were simulated with all
the six ($\sign$,$\alpha$) pairs of values; they are our fiducial halos.

%=======================

\section{Halo Structure}

%=======================

The six panels of Figure 3 show the density profiles of the fiducial 
cluster-sized halo ($Cl1$, solid lines) and MW-sized halo ($G1$, long-dashed 
lines) for the different ($\sign,\alpha$) pair values considered here. 
The fiducial halos $G1$ and $Cl1$ were selected in the low-resolution 
CDM simulations with concentrations typical for their masses in an
attempt to simulate halos with typical MAHs, since it is known 
that the concentration of the halo is related to its MAH  
(Avila-Reese et al. 1998, 1999; Wechsler et al. 2001). Unfortunately, in the 
high-resolution CDM simulation, the halos turned out to be more concentrated
(c$_{\rm NFW}=10$ instead of 6, and c$_{\rm NFW}=17.2$ instead of 13,
respectively). Besides, a major merger in the halo $Cl1$ that occurs at $a < 1$
in the low-resolution CDM simulation is delayed to $a \gtrsim 1$ in the 
corresponding high-resolution simulation; this will shift the core collapse 
phase to an epoch $a > 1$ when $\sig=3.0\ \cmgr$, as 
we will discuss below. In Figure 3, for 
reference, the density profiles of the two CDM models shown in panel (a) are 
also plotted with dotted lines in the other panels. The best NFW fits to 
the CDM profiles are shown with short-dashed lines in panel (a). 

From a first inspection of Figure 3, we conclude that significant differences 
in the density profiles (flattening) among the halos with different values of 
($\sign,\alpha$) apply only to those regions with radii smaller than 
$0.03-0.05$ \rvir. The exception are the halos 
with ($\sign,\alpha$) = (3.0,0.0); in this case SI affects dramatically the 
inner structure at radii even larger than $0.05 \rvir$. 
The parameter which defines the evolution under SI is 
the number of collisions per particle after a given time $t$, 
\begin{equation}
\ncol\propto \rho \left< \sig\vert\vrel\vert \right>t \propto \rho 
\sig \sigma_{3D}t,
\end{equation}
where $\sigma_{3D}$ is the 3D velocity dispersion.
The maximum  $\sigma_{3D}$ is proportional to \vmax. From all of our runs
we see that if \ncol\ at the present epoch is smaller than $\approx 2-5$ 
(see Table 1), then the halo is either in the core expansion phase or
has just had the gravothermal catastrophe triggered, being the core 
shrinkage still negligible. When \sig\ is constant, the cores of low 
velocity halos are on average less influenced by SI than the high 
velocity ones. On the other hand, 
when $\sig\propto 1/\vrel$, we expect halos of different sizes to
have roughly similar core densities. These predictions are closely obeyed 
as seen in panels  (b), (c), and panels (e) and (f) of Figure 3, respectively, 
or in Figure 4 below. Nevertheless, this simple reasoning applies strictly 
only for monolithic halos. The cosmological merger history may dramatically 
influences the $z=0$ structure of halos with significant SI. This is the case 
of the cluster-sized halo in the simulation with $\sig=3.0\ \cmgr$, 
\clfive\ (panel d). 
This halo should have evolved faster under the influence of SI than the 
MW-sized halo \Gnine. However, this did not happen; while halo \Gnine\ is
in an advanced stage of core collapse, halo \clfive\ is just entering into this
phase. The core collapse in the latter case seems to have been delayed by a
recent major merger. We have simulated a second cluster-sized
halo with \sig = 3.0 \cmgr, \clsix, which does not suffer late major mergers.
The core collapse phase at $a=1$ in this halo is well advanced and its 
central density and concentration are higher than those of the corresponding 
MW-sized halo (see Fig. 4 below). 

An important effect to note is that the core evolves fast after it
reaches its maximum expansion. The central density as well as \ncol\
(see eq. 4) increase quickly. 
This is why \ncol\ at $z = 0$ for the halo
\Gnine\ is similar to that of the halo \clfive (see Table 1) despite
that the latter has a \vmax\ 5.5 times larger than the former.
We recall that halo \clfive delayed its core collapse due to a late 
major merger.
Owing to the transient nature of the SIDM halo cores, a comparison of core 
parameters at a given epoch (e.g., $z = 0$) for different values of  
\sig\ and different halos masses and MAHs may lead to misleading conclusions.

In the following, we analyze in more detail the density profiles of the 
cluster and MW-sized SIDM halos in the light of previous works and then
explore the influence of the MAH on their evolution.

\subsubsection{Cluster-sized halos}

As mentioned above, it turned out that the chosen cluster-sized halo, which 
in the low-resolution CDM simulation was relaxed at $a = 1.0$ ($a = 1/(1+z)$), 
is in the high-resolution simulation about to suffer a major merger. The 
time when this merger happens seems to be a function of \sig. We find that 
the halo $Cl1$ with $(\sign,\alpha)=(3.0,0.0)$ suffers this merger before it 
reaches the present time, while $Cl1$ halos with other \sig\ values suffer this 
merger soon after $a = 1.0$. This is why in Figure 3 we compare the density 
profiles of halos $Cl1$ at time $a = 1.1$, an epoch when the $Cl1$ halos with 
different values of \sig\ have already experienced their last major merger.

We find the same general trend observed by Yoshida et al. (2000b) at $z = 0$: 
as \sig\ increases, the core radius, \rcsl, defined in this paper as the radius 
where density profile slope becomes steeper than $-1$, increases and the central 
density, \rhosl, decreases. However, differences between Yoshida et al. results 
and ours do exist: in our simulations there is a \sig\ (for the constant 
\sig\ case) for which the trend is reversed. For example, \rcsl\ for halo 
\clfive\ is not greater than the corresponding \rcsl\ of halo \clfour, 
and for halo \clsix, \rcsl\ is even much smaller. The trend of the core 
radius or central density with $\sig$ at $z=0$ is reversed owing to the 
core collapse. The halo with $\sig = 10.0\ \cmgr$ of Yoshida et al. is also 
undergoing a core collapse; however, at $z=0$ the central density of this 
halo is still larger than the central density of their halo simulated with 
$\sig = 1.0\ \cmgr$. Is the rate at which the core shrinks and the central 
density grows up in the stage of core collapse slower in their case than in 
ours?  We have measured this rate in both their cluster and our cluster halo 
\clfive\ and have found that the rates are actually similar. The difference 
between ours and Yoshida's et al. results may be found in the minimum central 
density reached by the halos. In their case, this density is lower than in 
our case by a factor of three. This is probably why the soft core at $z = 0$ 
of their cluster halo simulated with $\sig = 10.0\ \cmgr$ have not become 
yet smaller than the corresponding one to the simulation with  
$\sig = 1.0\ \cmgr$, in contrast to our results where for $\sig = 3.0\ \cmgr$  
the soft core is already smaller than for  $\sig = 0.5\ \cmgr$. 

Since halo \clfive\ assembles most of its mass through late major mergers, 
the triggering of the core collapse phase, which is almost inevitably 
present in almost all halos with ($\sign,\alpha$) = (3.0,0.0), 
is delayed, and at $a = 1.1$ still has a prominent soft core, 
although it is already shrinking because of the
core collapse. After the last major merger took place, 
little mass is accreted by the halo in the next $3.5 \times 10^9$ yr; 
it is thus tempting to compare the core evolution since that
epoch with the predictions of the monolithic case of KW. 
The halo \clfive\ would have a \siga\ that lies in 
the interval [3.5,4.0]\footnote{Aside from \sign\
we also need to know the concentration and the total mass in
order to compute \siga. We have taken these parameters from its
correspondent CDM cluster halo. This is a good approximation because
once the cluster forms, which we can establish at the last major merger,
the parameter $r_{NFW}$ does not change much.} and a dynamical time 
\tdy, as defined by KW, of
$1.7 \times 10^9$ yr. From $a = 0.85$, which is the epoch where 
the maximum core expansion occurs, to $a = 1.10$ we see that 
the core shrinks by a factor of 1.3. KW predict for this model
a shrinking factor of about 1.4-1.5. It is not clear how
far we can take the comparison of the SIDM cosmological scenario with
the monolithic case, but here we have a case in which consistent results 
are obtained.
{\pspicture(-0.6,-0.5)(12.0,16.5)
\rput[tl]{0}(-0.5,16.0){\epsfxsize=8.5cm
\epsffile{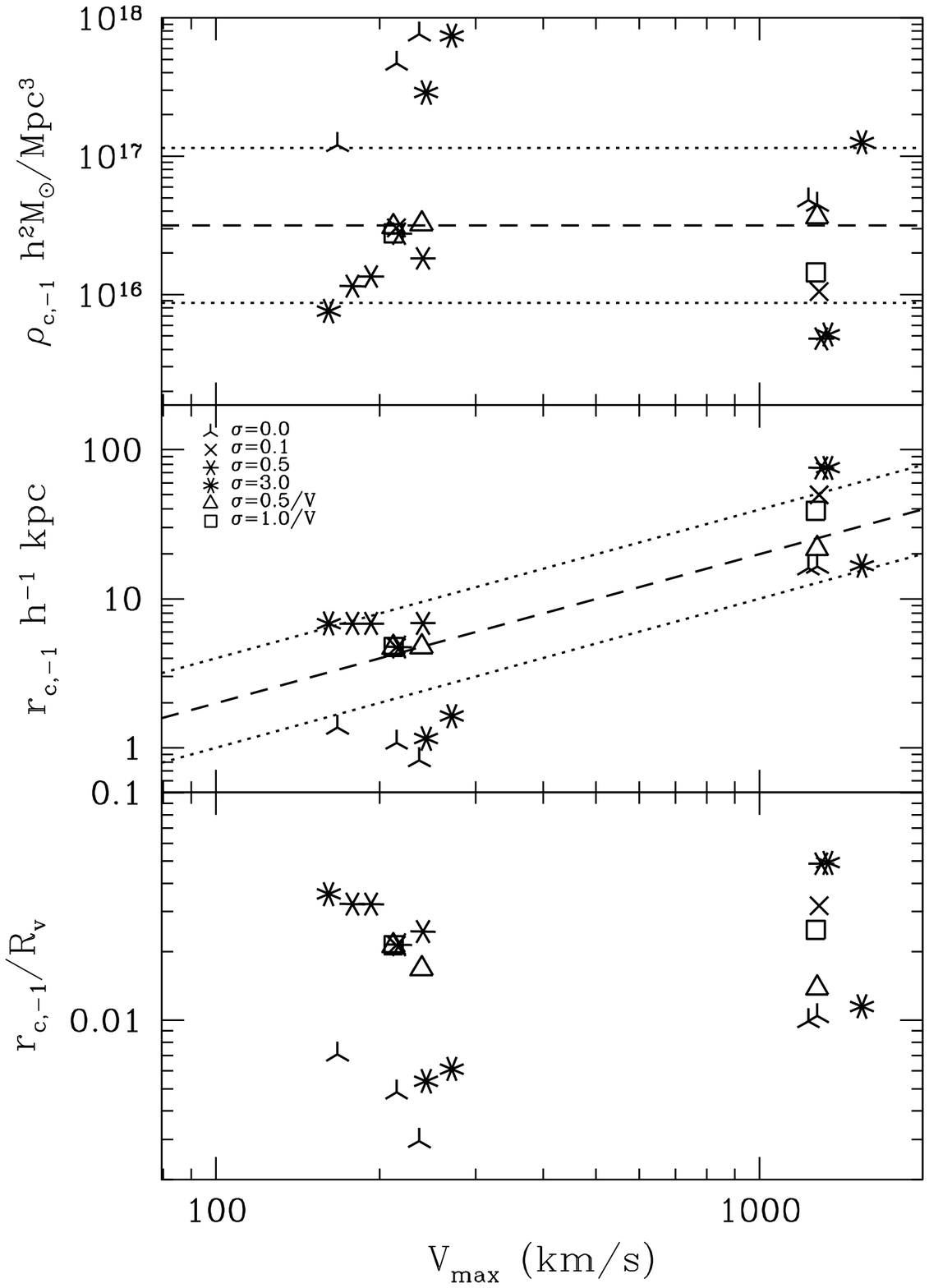}}
\rput[tl]{0}(-0.7,3.7){
\begin{minipage}{8.8cm}
  \small\parindent=3.5mm {\sc Fig.}~4.---
Central density, core radius and core fraction vs.  
\vmax\ for 
cluster- and MW-sized halos simulated here for different values of 
\sig\ (see the corresponding symbols in the middle panel; \sig\ is in units of 
\cmgr\ and V is \vrel\ in units of 100 \kms). Central density \rhosl\ and core 
radius \rcsl\ are deffined as the density and radius where the profile slope 
becomes steeper than $-1$, respectively. Dashed and dotted lines in the two upper 
panels are the average and the $2-\sigma$ dispersion inferred from 
observations of dwarf and LSB galaxies and cluster of galaxies (see $\S 5.2$ for 
details).
\end{minipage}}
\endpspicture}

\subsubsection{Milky Way-sized halo $G1$}

The fiducial MW-sized halo $G1$ resulted more concentrated 
(see above) and with a MAH more biased to an early halo-assembly than 
the average corresponding to its mass. Moreover, the late MAH of this halo 
is nearly smooth, without violent major mergers since $a = 0.5$. 

The non-monothonic trend of the core radius or the central
density with \sig\ at $z=0$ sliglthy outlined by the fiducial cluster-sized halo
is clearly present in the fiducial MW-sized halo (Fig. 3).
This is much better appreciated in Figure 4. 
Here we plot \vmax\ versus central density, \rhosl, measured at the radius 
\rcsl\ where the profile slope becomes steeper than $-1$ (a), the core 
radius, \rcsl\, (b), and the core radius fraction, \rcsl/\rvir, (c). Note
that halos with ($\sign,\alpha$) = (0.1,0.0) and (0.5,0.0) 
tend to lie in the same region; \ie, an increase in \sign\ does not
necessarily produce an increase in core radius at $z=0$. In the framework 
of a monolothic halo this could be explained as follows: because the core 
relaxation time for halos with ($\sign,\alpha$) = (0.1,0.0) is close or 
greater than the Hubble time, we still find these halos in 
their core expansion phase at the present time, whereas halos with
($\sign,\alpha$) = (0.5,0.0) are at the onset
of the core collapse phase. These facts conspire to give
similar core parameters in both cases. Halo \Gnine\ on the other 
hand, is well inside the core collapse phase and this is why
\rcsl\ is small. It is interesting to note that SI 
introduces a peculiar epoch in the history of the universe at which
the dark halos change their inner structure from an extended smooth 
core to a very cuspy center. For the case $\sig\propto 1/\vrel$, 
this epoch does not depend too much on the halo mass or \vmax, and it is 
determined mainly by the value of $\sign$. However, violent mergers and 
other effects may delay significantly the core collapse phase for some halos 
and subhalos (see \S 3.1.3).

In Figures 5 and 6 we plot the evolution of the density and 3D velocity dispersion
profiles, and of the heat capacity profile, respectively, of the core collapsing 
halo \Gnine\ from $a = 0.4$ to $a = 1.0$ (left panels). Proper units were used.
As in the case of halo \clfive, one could ask 
whether the strong evolution of \rhoc\ is expected or not. To address this 
question we first notice that only 20\% of the present total mass is
accreted since the last major merger, which occurs at $a \simeq 0.5$. 
Therefore, treating the 
evolution of this halo from $a = 0.5$ to $a = 1$ as an isolated case seems to be
a fair approximation\footnote{We later realized that what matters is not only
{\it the amount of mass} but {\it how this mass} is accreted. Smooth infall
of material to halo outskirts does not seem to alter significantly the dynamical 
core SI evolution}. As we did for the fiducial cluster-sized halo, 
we compute the corresponding \siga\ and dynamical time; they
are 1.9 and $6.4 \times 10^8$ yr, respectively. From the KW simulations of 
monolithic halos, this model changes its core radius by a factor of four in 
approximately 5-6\tdy; this same change is seen in halo \Gnine\ but
in about 10\tdy. We can explain this difference in the core evolution rate 
as being due to the infalling and merging material in the cosmological halo.
This also could explain why the inner region where the heat capacity $C$ is
positive increases during the expansion phase in halo \Gnine\ (Fig. 6) unlike
what happens with the monolithic halo in \S 2.1 (Fig. 1). The subsequent core 
collapse phase is characterized by a reduction of the region where $C > 0$. 
\begin{figure*}[ht]
\pspicture(0.5,-1.0)(15.0,12.0)
\rput[tl]{0}(1.0,11.5){\epsfxsize=18cm
\epsffile{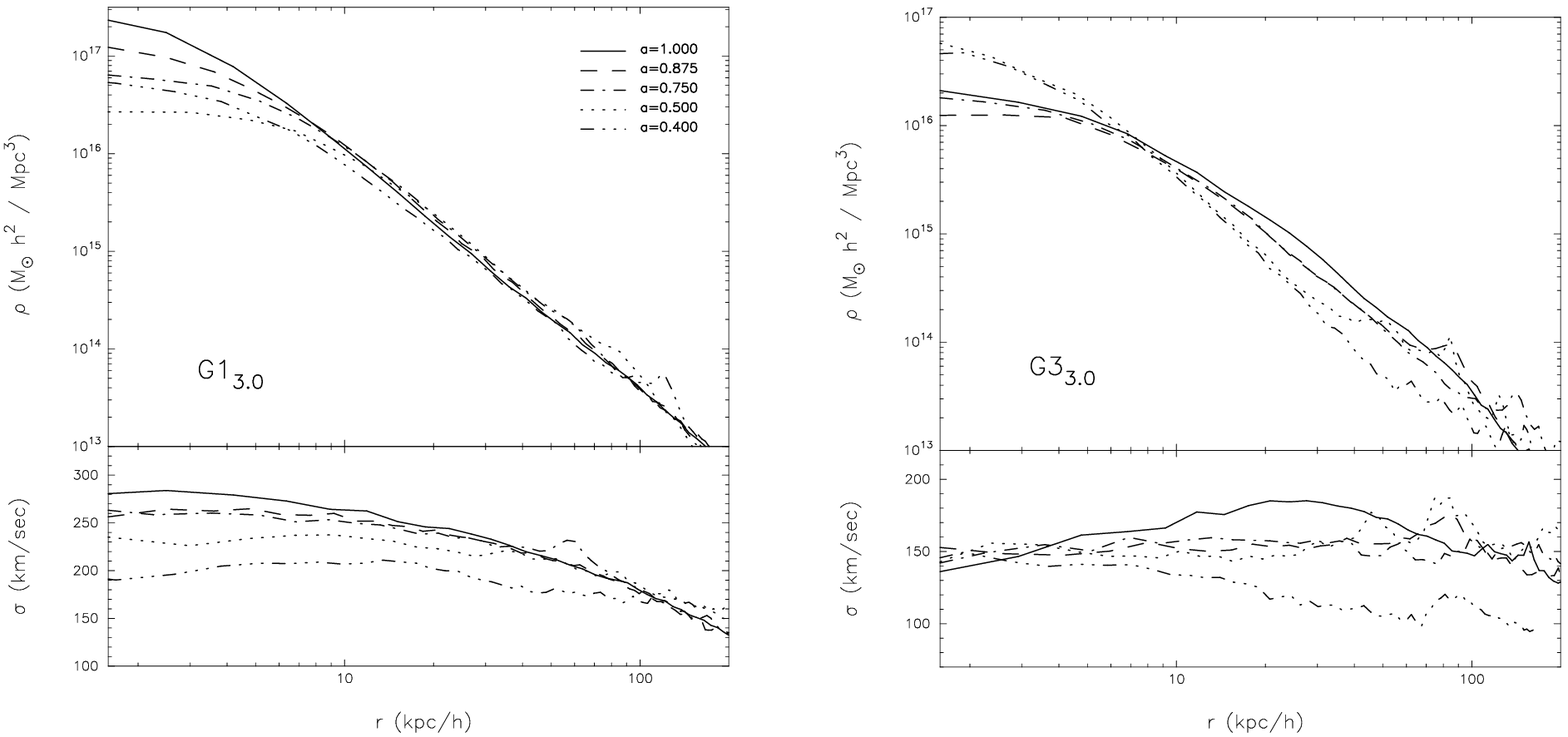}}
\rput[tl]{0}(0.5,1.8){
\begin{minipage}{18.4cm}
  \small\parindent=3.5mm {\sc Fig.}~5.---
Evolution of the density and the 3D velocity dispersion 
profiles of the $G1$ (left) and $G3$ (right) halos in proper units and 
for $(\sig,\alpha) = (3.0,0.0)$. In both halos the core reaches its
maximum size a little later after the major marger occurs, around $a = 0.5$
for \Gnine\ and $a \simeq 1.0$ for \Geleven; 
for times earlier than this, the central densities of the major progenitors of  
the halos are higher. The shape of the density profiles of their corresponding 
CDM halos do not change after the major mergers occur, the subsequent infall of 
material serves essentially to increase the proper size of the halo. 
Because a very small amount of material reach the core (see also the paper by 
\cite{Zhao02}), we can safely say that the inner regions of the SI halos
evolved as if they were isolated; in other words, the number of collisions inside
the core overwhelms the number of particles that fall into it. So, it is not
very surprising, after all, that we also see a core collapse in MW-sized {\it 
cosmological} halos.
\end{minipage}}
\endpspicture
\end{figure*}

The core collapse of halo \Gnine\ (and \Gten, see below) is at odds with 
the results of Dav\'e et al. (2001). They ran a simulation of $128^3$ 
particles in a 4 \mpch\ box for the same cosmological parameters used here. 
Their power spectrum is normalized to $\sige = 0.8$. The most 
massive halo formed in their simulation has a mass of 
$6.2 \times 10^{11}\ \msun$ and consists of about  $1.7 \times 10^5$ 
particles. None of the halos in the two SI models that they consider, (0.56,0.0) 
and (5.6,0.0), shows signs of being in the core collapse phase. On the contrary, 
the halos have extended shallow cores which increase as \sig\ 
increases. Dav\'e et al. suggested that the accretion of dynamically hot 
material, intrinsic to the CDM cosmological simulations, is so important 
to the halo core evolution that predictions for SIDM from monolithic halos 
does not apply at all to the cosmological setting.

\subsubsection{Halos with different assembling histories}

In the following, we test the posibility that our differences with 
Dav\'e et al. (2001) might reside in the halo MAHs. It is well known that
CDM halos form from a variety of MAHs which give rise to a variety
of present-day structures (e.g., Avila-Reese et al. 1998,1999). As 
mentioned above, the fiducial halo assembled early and ended highly 
concentrated. We need to 
simulate now a halo with a low concentration and a late halo-assembly.
We chose two MW-sized halos with these characteristics, \Gtwo\ and \Gthree, 
from the same low-mass resolution simulation as the one from which the 
halo \Gone\ was taken; \Gten\ and \Geleven\ are the corresponding SI halos 
with ($\sign,\alpha$) = (3.0,0.0). As for the fiducial halo, in the high 
resolution simulation the halos 
resulted with a higher concentration and a less extended MAH than in the
low resolution one. In any case, both halos are less concentrated and have 
a more extended MAH than the fiducial one. 

To our surprise, halo \Gten\ also undergoes a fast core collapse, no 
matter that this halo accretes a little more than half of its present 
mass from $a = 0.5$ to $a = 1.0$!  An analysis of the MAH of the 
corresponding CDM halo shows that the last major merger occurs as early 
as $a \simeq 0.3$; after this epoch the mass is accreted {\it smoothly} 
and/or in small chunks. A recent paper by Zhao et al. (2002) shows that
the build-up of CDM halos can be divided in time in two phases: an early one 
characterized by violent mergers in which the inner halo density profile
is significantly altered, and a late one characterized by a smooth acretion
of material in which the accretted material retains its low binding energy and 
is added to the outer part of the halo. Depending on the MAH, a halo can
go from one phase to the other early or late in time. Halos \Gone\ and \Gtwo\
appear to have typical transition epochs, although the latter accretes 
(smoothly) much more mass at late epochs than the former one. On the other 
hand, halo \Gthree\ seems to have the transition at $z \simeq 0$. This halo 
assembles most of its mass by recent major mergers. 
\begin{figure*}[ht]
\pspicture(0.5,-1.0)(15.0,10.5)
\rput[tl]{0}(1.0,10.0){\epsfxsize=18cm
\epsffile{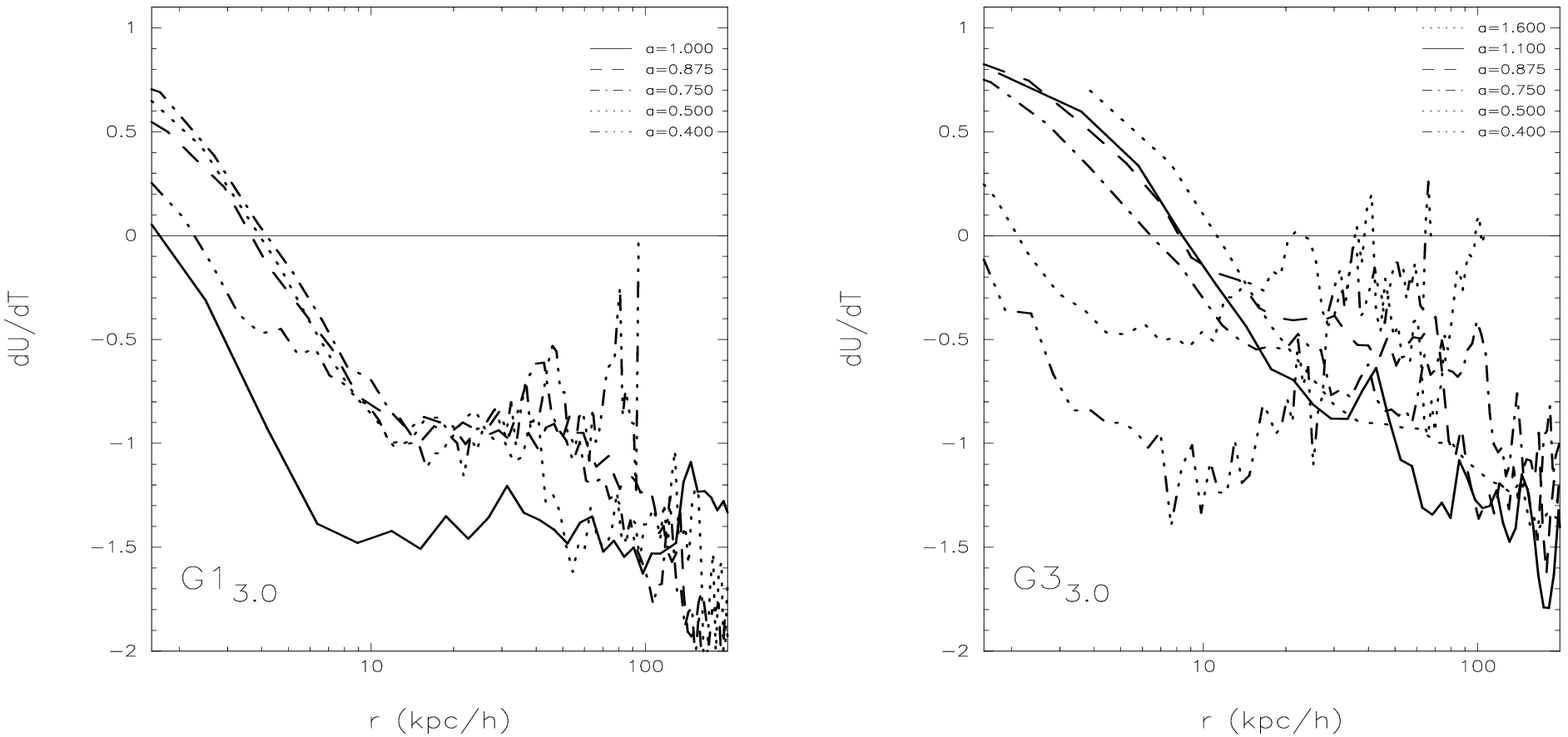}}
\rput[tl]{0}(0.5,1.0){
\begin{minipage}{18.4cm}
  \small\parindent=3.5mm {\sc Fig.}~6.---
Evolution of the heat capacity, $C \equiv dU/dT$, 
profiles for halos \Gnine\ (left panel) and \Geleven\ (right panel). Notice 
how the size of the region where heat capacity is positive increases 
in both halos as the cores of the halos approach their maximum radius. 
As the core collapse proceeds in halo \Gnine\ the region where $C$ is 
positive shrinks. This is different to what we see in the subsequent 
evolution of $C$ in halo \Gnine, where the core collapse is absent; 
in this case the size of the $C>0$ region even increases as time passes.
\end{minipage}}
\endpspicture
\end{figure*}

Besides of the direct dynamical effect of the mergers on the structure of
halo \Geleven, its core collapse may be delayed by conduction effects.
The major mergers heat the halo periphery, producing an increment of the
outer velocity dispersion (temperature), as it seen in the right panel of 
Fig. 5 (dotted line). In this case, the halo at $a=0.4$ was already starting 
its core collapse phase. The heat capacity also increases due to the major 
mergers (Fig. 6). Then, heat starts flowing into the $C>0$ core and produce 
its expansion (see \S 2.1). 
This process may last a significant period of time, until the heat conduction 
isothermalise most of the halo. We let the simulation of halo \Geleven\ run 
for another Hubble time and to our surprise even at $a = 1.6$, about 20 Gyr 
after the Big Bang, the halo has not undergone a core collapse and the heat 
capacity increased even slighlty. We should note that this halo is also peculiar 
because close to it there is a galaxy-sized more massive halo. We notice, on 
the other hand, that because halo \Geleven\ is smaller ($\vmax=163$ \kms) 
than halos \Gnine\ and \Gten, we expect the SI evolution of this halo to
be slower than that of halos \Gnine\ and \Gten.

Our results show that, for both the cluster- and MW-sized halos
with ($\sig,\alpha)=(3.0,0.0)$, the core collapse is delayed until the stage 
of {\it major mergers} ends. We further simulated the halo G1 with a 
still higher \sig\ ($= 10.0$ \cmgr) and check, in fact, that the core collapse 
begins only after the last major merger. Therefore, we conclude that the dynamics 
of SIDM halos with a high cross section is strongly affected by {\it 
violent merging processes rather than by smooth mass aggregation}. A 
continuous injection of ``dynamical heat'', able to either heat and expand 
directly the core or rise the temperature of the halo periphery to above the 
core's temperature, seems to be possible only when the halo grows by violent 
mergers.  In summary, when SI is efficient ($\sig = 3.0\  \cmgr$, for example) 
the inner structure of halos may depend strongly on the halo merging history. 
The onset of the core collapse phase occurs immediately after the core relaxes, 
in agreement with the finding of KW (see also Quinlann 1996; Burkert 2000), 
but this process can be delayed by subsequent major mergers as we actually see 
in some of our halos. Can this explain our differences with Dav\'e et al. (2001) 
results?

\subsection{Concentrations}

A more quantitative way to measure differences in the structure among 
halos is using concentration parameters. In Figure 7 we plot the 
c$_{1/5}$ and c$_M$ concentration parameters\footnote{
c$_{1/5}$ is defined as the ratio between the virial radius \rvir\
and the radius where 1/5 of the total halo mass is contained 
(Avila-Reese et al. 1999). c$_M$ is defined as 27 times the ratio 
between the mass at r$_{\rm in}=8.5$ kpc$\ \vmax/220 \kms$ and the mass
at r$_{\rm out}=3\times$r$_{\rm in}$ (Dave et al. 2001). These definitions of 
the concentration parameter are independent from the particular fitting 
applied to the halo density profile.} versus virial mass \mvir\ for 
halos (large symbols) and subhalos with more than 1000 particles 
(small symbols) from our different simulations. To avoid too much overlapping,
we divided Figure 7 in two sequences of panels. Halos with
$(\sign,\alpha) = (0.0,0.0)$, (0.1,0.0), (3.0,0.0),
and (1.0,1.0) are shown in left panels while halos with
$(\sign,\alpha) = (0.5,0.0)$ and (0.5,1.0) are shown in right panels.
We also plot in Figure 7 the density measured at $0.03\rvir$ 
($\rho_{\rm c,.03}$) versus \mvir. Because our high-resolution
simulations were designed to be focused on a specific (cluster- or MW-sized)
halo, most subhalos lie within the virial radii of these halos (satellites). 
The mass of subhalos is defined as the minimum between the virial mass and 
the truncated mass (see Avila-Reese et al. 1999). 

From Figure 7 one sees that halos with different $(\sign,\alpha)$ values 
have similar c$_{1/5}$ parameters, except for halos \Gnine\ and
\Gten, in which the region where 1/5th of \mvir\ is contained is
greatly affected by the core collapse, and halo \Geleven, whose
peculiar MAH makes it keep a relatively long-lived shallow 
core (see \S 3.1.3).
A similar behavior is observed for $\rho_{\rm c,.03}$. The parameter 
c$_M$ reflects with better accuracy the inner matter distribution of 
halos. Note that the c$_M$ parameter of subhalos with more than 1000 
particles from the MW-sized $(\sign,\alpha) = (0.5,0.0)$ simulations are 
similar to those from the $(\sign,\alpha) = (0.56,0.0)$ simulation of 
\cite{Dave2001}. Halos with higher \sign\ have smaller c$_M$ concentrations, 
for a given $\alpha$. Again, this trend can be reversed at the high \sign\
end because of the core collapse. Subhalos follow roughly the same behavior 
of halos in all the three parameters of Figure 7 as a function of mass. However, 
before drawing any conclusion we should bear in mind that (i) subhalos are 
resolved only with hundreds of particles and (ii) that they can result strongly
affected by the self-interacting environment of the halo where they are embedded.

The lack of mass resolution does not allow us to explore in detail the effect 
of SI on the inner dynamics and structure of subhalos. 
In order to evaluate this problem, we analyze the MW-sized halos obtained
in a low resolution ($64^3$ particles) simulation with  
$(\sign,\alpha) = (3.0,0.0)$. This is the same simulation from where 
MW-sized halos were taken. The SIDM MW-sized
halos in this simulation have only few thousands of particles and we find
that their inner structures are not at all similar to those of their 
high-resolution counterparts. Therefore, we can not be conclusive about the
inner structures of the subhalos in the high-resolution simulations presented 
here because they have at the most only a few thousands of particles. 
Nevertheless, our simulations allow us to explore the effect the hot 
particles of the host halo has on the overall structure of subhalos. 
The c$_{1/5}$, c$_M$, and $\rho_{\rm c,.03}$ parameters of
the subhalos from the cluster-sized halo simulation \clfive\
are smaller than those same parameters of the isolated MW-sized 
halos from the low resolution simulation. In both cases the halo masses
and number of particles are similar. This difference seems to have 
only one explanation: the structure of subhalos (satellites) is 
affected by the environment of the hot host halo. Subhalo particles 
are being ejected continuously because high speed particles 
from the hot host halo collide with the cooler subhalo 
particles, exchanging orbital energy by internal energy; in the extreme case, 
the subhalo can be evaporated completely (Gnedin \& Ostriker 2002).  The 
survivor subhalos are indeed ``puffy'' with density profiles much flatter 
than the isolated halos.

More evidence about the influence of SIDM halos on the structure of their 
subhalos comes from comparing subhalos of simulation \Gfourteen\ 
with subhalos of simulation \Gnine\ (Fig. 7).
Subhalos from the former simulation (squares) have an effective \sig\ of 
$\approx 3-5$ \cmgr, \ie, similar to the \sig\ of the latter simulation (lowest 
starred symbols); however, they are located in diferent regions in Figure 7 
(the same happens for subhalos with less than 1000 particles). The difference 
between both simulations is in the effective \sig\ value of the host halos: 
$\sim 0.5 \ \cmgr$  versus $3\ \cmgr$, respectively. 
The evaporation effect of the hot SIDM halo \Gnine\ is stronger
than the one of the halo \Gfourteen; in fact, in the former case most 
subhalos have been destroyed by the present time (see \S 4).  
A similar effect is observed in
subhalos from the $(\sign,\alpha) = (0.5,1.0)$ and (0.5,0.0) simulations.
For the MW-sized halos of the former simulation, the effective value of 
\sig\ is $\approx 0.2\ \cmgr$ while for their subhalos it is 
$\sig \approx 1.5-3.0\ \cmgr$; \ie, 3-6 times larger than for subhalos 
of the latter simulation. Despite the relatively high value of \sig\ of 
subhalos in the (0.5,1.0) model, their
structures look similar to their CDM counterparts (low-mass resolution
does not allow SI to work properly). However, subhalos from the 
simulation with $\sig = 0.5\ \cmgr$ are ``puffier'' than those subhalos from
CDM (right panels, Fig. 7). This can also be explained by
an action of the hot halo environment on the dynamics of subhalos.
{\pspicture(-0.6,2.0)(12.0,18.5)
\rput[tl]{0}(-0.5,18.0){\epsfxsize=8.5cm
\epsffile{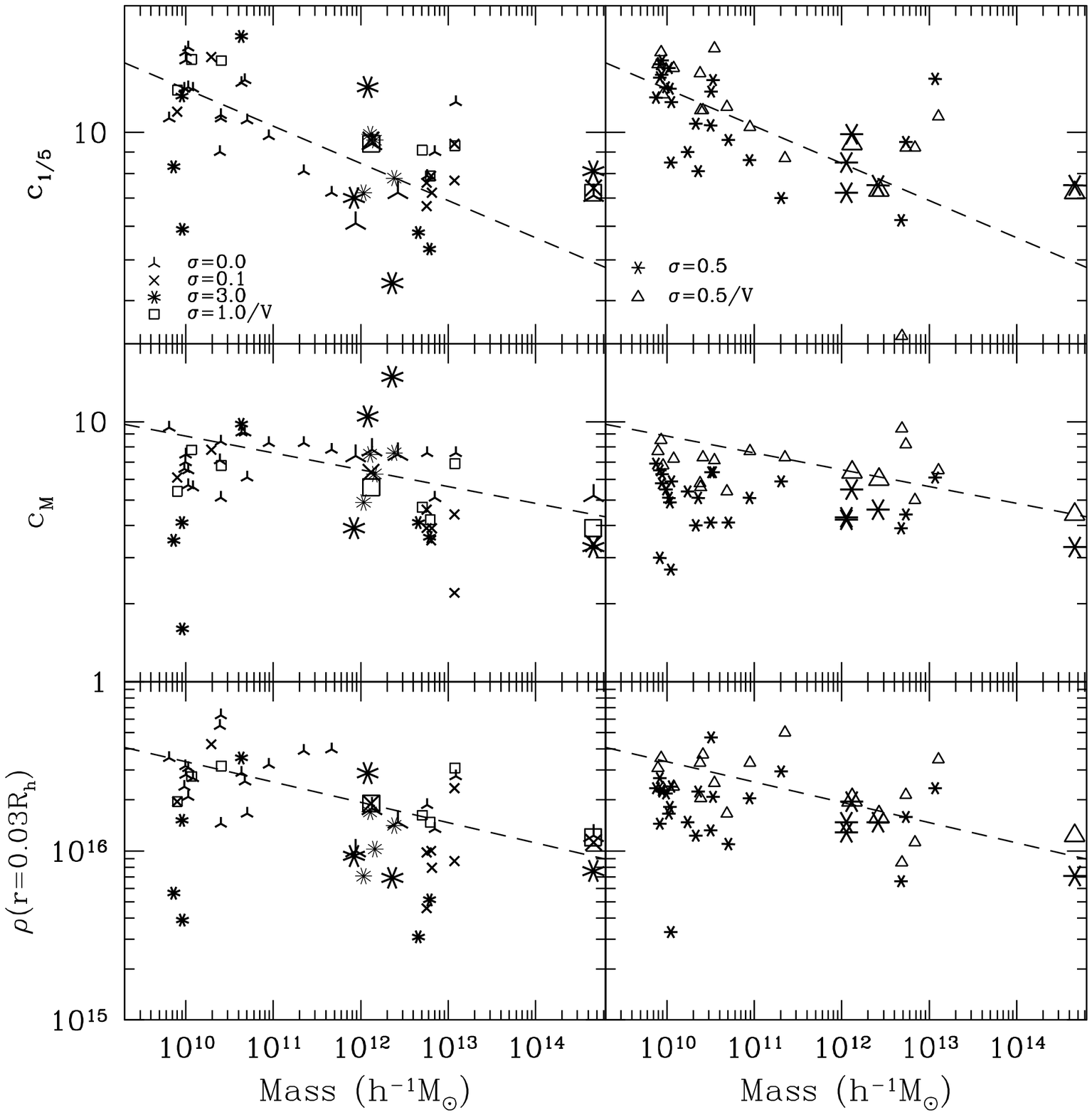}}
\rput[tl]{0}(-0.7,8.7){
\begin{minipage}{8.8cm}
  \small\parindent=3.5mm {\sc Fig.}~7.---
Concentration parameters c$_{1/5}$ and c$_M$, and density 
$\rho_{\rm c,.03}$ at 0.03\rvir\ versus halo mass for halos (large 
symbols) and guest halos with more than 1000 particles (small symbols) 
from simulations with different values of \sig\ (see the corresponding symbols 
in upper panels; same unities as in Fig. 3). Results were 
divided in two sequences of panels in order to avoid too much overlapping.
Tiny starred symbols in the left panels are for the low-mass resolution
simulation in the 12.5 \mpch\ box with $(\sign,\alpha) = (3.0,0.0)$. Their
structural properties are different to that of their high-mass resolution
counterparts (bright starred symbols). Dashed line in the upper panels is
the linear regression found in Avila-Reese et al. (1999) for isolated $\Lambda$CDM
halos in a 60 \mpch\ box. Dashed line in medium and lower panels are estimates
of the c$_M-\mvir$ and $\rho_{\rm c,.03}-\mvir$ relations for isolated 
$\Lambda$CDM halos from the the same simulation of  Avila-Reese et al. (1999).
c$_M$ was calculated from c$_{\rm NFW}$ using Fig. 6 in Dav\'e et al. (2001).
\end{minipage}}
\endpspicture}

\subsection{Ellipticities}

Another prediction by the SIDM cosmology is that the halo inner region, where a
shallow core forms, should be close to spherical because collisions tend to reduce
the degree of anisotropy. 
\begin{figure*}[ht]
\pspicture(0.5,-1.0)(15.0,9.2)
\rput[tl]{0}(1.0,8.7){\epsfxsize=18cm
\epsffile{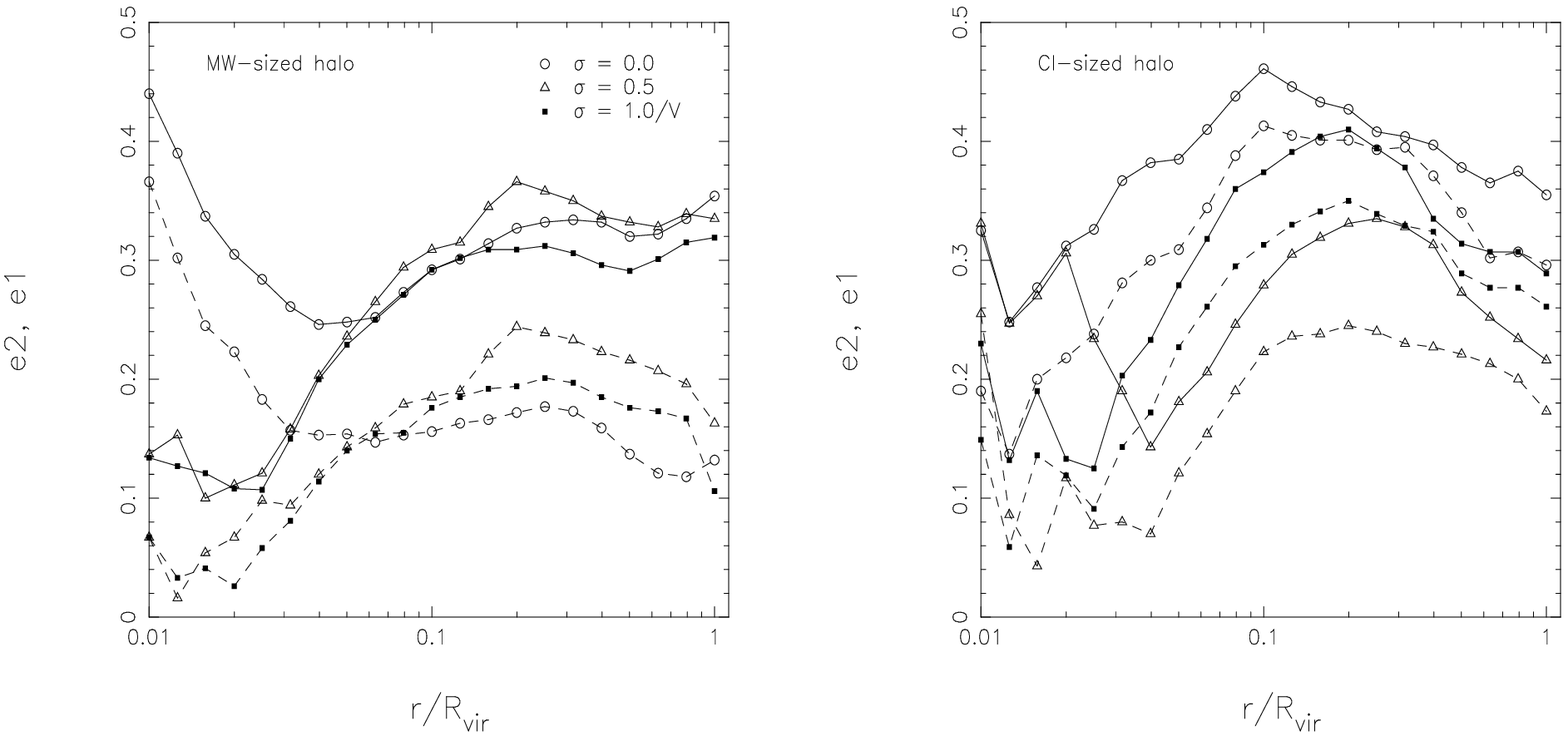}}
\rput[tl]{0}(0.5,-0.3){
\begin{minipage}{18.4cm}
  \small\parindent=3.5mm {\sc Fig.}~8.---
Ellipticities $e1$ and $e2$ as a function of radius 
are shown in left (halo $G1$) and right (halo $Cl1$) panels. As expected, SI halo
are rounder than the corresponding CDM halos, being the effect greater for 
the cluster-sized halo.
\end{minipage}}
\endpspicture
\end{figure*}

We have measured the ellipticities of halos using the tensor of
inertia. This is defined as
\begin{equation}
I_{i,j} = \sum x_i x_j / r^2,
\end{equation}
where the sum is over all particles within $r_{vir}$, $x_i$ ($i=1,2,3$)
are the particle coordinates with respect to the halo center of mass,
and $r$ is the distance of the particle to the halo center.
The ellipticities are then given by
\begin{equation}
e_1 = 1 - \frac{\lambda_1}{\lambda_3},\ \ \ \ \ \ \ \ \ \ \  \ e_2 = 
1 - \frac{\lambda_2}{\lambda_3},
\end{equation}
where $\lambda_3 > \lambda_2 > \lambda_1$ are the eigenvalues of the tensor
of inertia. We evaluate the triaxiality parameter using the following
formula (\eg \cite{FIdZ90})
\begin{equation}
T = \frac{\lambda_3^2 - \lambda_2^2}{\lambda_3^2 - \lambda_1^2}.
\end{equation}
A halo is prolate (oblate) if $T = 1.0$ ($T = 0.0$).

The ellipticities $e_1$ and $e_2$ as a function of radius are
shown in Figure 8 for our fiducial MW- and cluster-sized 
halos for three
models $\sig = 0.0$, 0.5, and $1.0/V \cmgr$. We have denoted with solid 
and dashed lines $e1$ and $e2$, respectively.  We see, as expected, that
the differences in the ellipticities between SI and CDM halos are found 
in the inner parts. At large radii, the difference in $e1$ or $e2$ values
among the plotted MW-sized halos is lower than the difference between the 
values of the ellipticities $e1$ and $e2$. The opposite occurs for the
cluster-sized halo. Note, moreover, that unlike $G1$ halos, SI $Cl1$ halos are 
rounder at all radii. 

Cores of SI halos are indeed rounder than their CDM counterparts, but 
they are not spherical. Unfortunately, ellipticities in real galaxies are 
ill determined, so it is not clear how much we can constrain the SI 
parameters with this kind of observations. Miralda-Escud\'e (2002) claims, 
however, to have constrained the constant cross section
to be less than $\simeq 0.02 \cmgr$ with the determination of the ellipticity
of cluster MS 2137-23 through a lensing study. The question remains whether the
ellipticity he derives for MS 2137-23 does disagree with the one found 
in a typical cluster-sized {\it simulated halo} with $\sig >  0.02 \cmgr$.
We let this analysis for a subsequent paper.

\subsection{Angular Momentum Distribution}

For each halo we compute the total angular momentum as
\begin{equation}
\mbox{\boldmath $J$} = m_i \sum_{i=1}^{n} \mbox{\boldmath $r_i$} 
\times \mbox{\boldmath $v_i$},
\end{equation}
where \mbox{\boldmath $r_i$} and \mbox{\boldmath $v_i$} are the position and 
velocity of the {\it i}th particle with respect to the halo
center of mass. We follow Bullock et al. (2001) and define
a modified spin parameter  $\lambda'$ to characterize the 
global angular momentum of a halo
\begin{equation} 
\lambda' \equiv \frac{J_{vir}}{\sqrt{2} M_{vir} V_{vir} R_{vir}},
\end{equation}
where $J_{vir}$ is the angular momentum inside the virial radius $R_{vir}$, 
and $V_{vir}$ is the circular velocity at radius $R_{vir}$. In Table 1, column
(8), we show $\lambda'$ for all of our halos. Notice that $\lambda'$ 
increases with \sig\ in both cluster- and MW-sized halos for constant 
\sig, from $\sign = 0.0$ to 0.5. Nevertheless, this effect is small, 
amounting to 10\%, 16\% and 3\% for halos $Cl1$, $G1$ and
$G2$, respectively. It measures roughly the difference in the number of 
particles with low angular momentum in inner regions between CDM
and SI halos. Moreover, for those halos for which
$\sig \propto 1/\vrel$, the effective \sig\ is so small that a factor of
two of change in \sign\ passes unnoticed by $\lambda'$.

These results already suggest that significant differences in the angular 
momentum distributions (AMDs) of halos with and without SI are not expected.
To compute the AMD we find first, for each particle, the specific angular 
momentum component along $\mbox{\boldmath $J_{vir}$}$, calling it $j$. 
We then divide the sphere of radius $R_{vir}$ in spherical shells and 
each of these shells is in turn divided in four quadrants. Differences
in the number of particles between cells for most of them are below
a factor of two. Only a small fraction of these cells
have negative $j$ values and we reject them in the further 
calculation of the AMD; notice, however, that these particles with negative
$j$ can have a significant effect on the angular momentum problem and the
formation of bulges (\cite{vandenBosch2002}). Cells are ranked according to
their $j$ value and $M(<j)$ profiles are then built by counting the cumulative
mass in cells with angular momentum smaller than $j$. Figure 9 shows the
$M(<j)$ profiles for halos $G1$ and $Cl1$ (panels a and b, respectively, where
circles are for CDM and squares for $\sig = 0.5$). Mass
and angular momentum are given in $M_{vir}$ and $j_{max}$ units, respectively, 
where $j_{max}$ is the maximum value reached by $j$. We 
have also plotted in Figure 9 the analytical form of
$M(< j)$ proposed by Bullock et al. (2001) with $\mu = 1.25$; the line
intends to guide the eye, it is not a fit to our numerical results. The
effect of the dynamical expansion of the core is seen only at low
$j$. There is less mass, for a given $j$, in SI halos or, for the same
fraction of mass, CDM halos have less rotational support. However,
the difference between the CDM and the SI profiles is so small that SIDM
can not account for the differences between the predicted angular
momentum distribution of the CDM model and that determined for dwarf
galaxies (\cite{vBS2001}).
\begin{figure*}[ht]
\pspicture(0.5,-1.0)(15.0,10.7)
\rput[tl]{0}(1.0,10.2){\epsfxsize=18cm
\epsffile{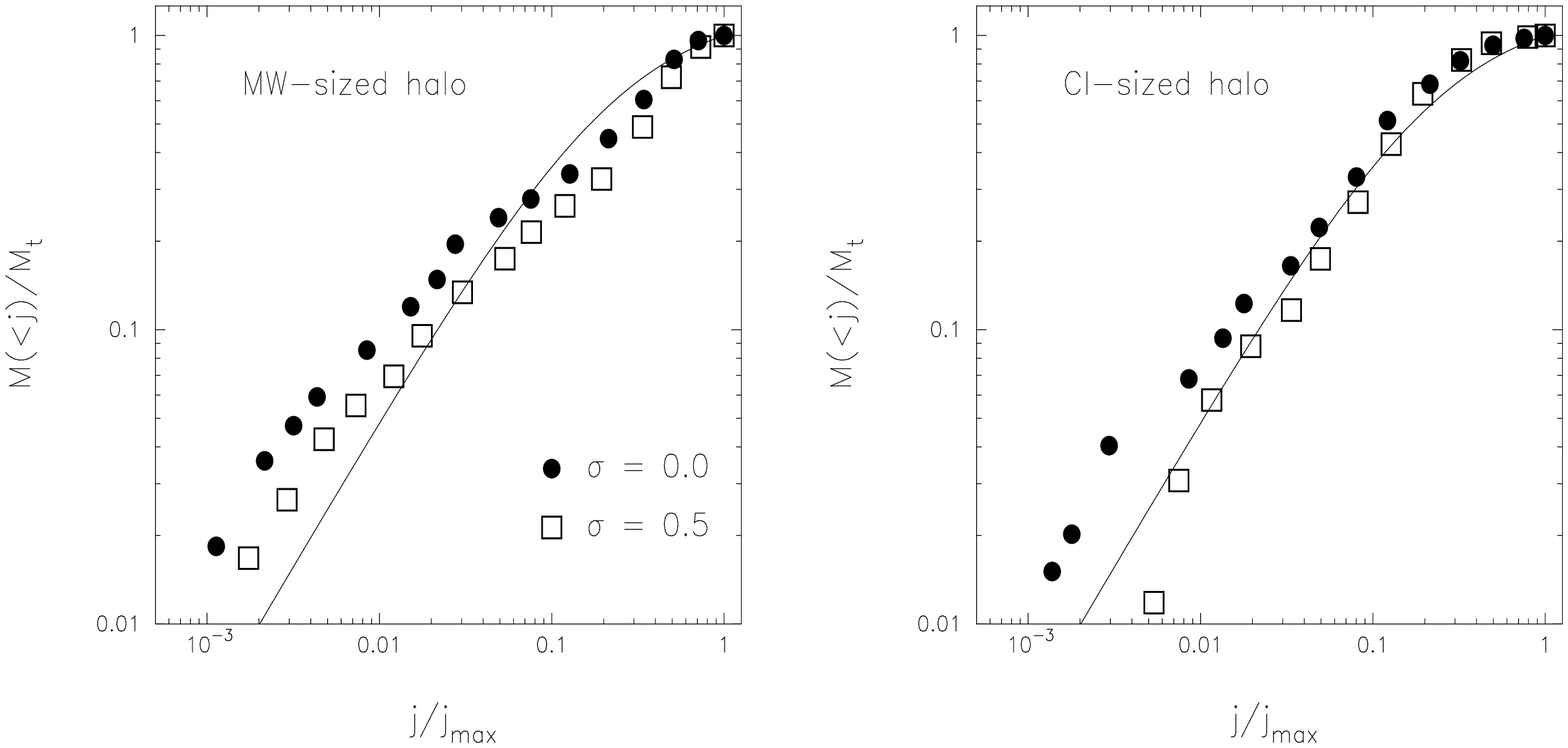}}
\rput[tl]{0}(0.5,1.2){
\begin{minipage}{18.4cm}
  \small\parindent=3.5mm {\sc Fig.}~9.---
Specific angular momentum $j$ distribution for halos 
\Gone\ (circles)
and \Gfive\ (squares) panel (a) and for halos \clone\ (circles) and \clfour\
(squares) panel (b). Only cells with positive $j$ were considered and the total
mass $M_t$ was redefined to be the mass contained in all those cells with
positive $j$. The line is taken from the analytical fit of Bullock et al. (2001)
$m(<j) = \mu \tilde{j} / 1 + \tilde{j}$, where $m(<j) \equiv M(<j)/M_{vir}$
and $\tilde{j} \equiv j/j_{max}$, with $\mu = 1.25$. Notice that differences
between the CDM and SI $m(<j)$ profiles are only at the low-$j$ end. We see that
our fiducial MW-sized halo is not well represented by the formula of Bullock
et al., which agree with the findings by \cite{ChJ2002}.
\end{minipage}}
\endpspicture
\end{figure*}

\section{Subhalo Population}

In previous works on SIDM it was suggested that the number of subhalos 
in MW-sized halos should be lower than that predicted in CDM because 
subhalos are supposed to form with a lower concentration and/or because, 
the hot environment of a SIDM halo evaporates its subhalos. What do we 
see in our simulations? For halos \Gnine, \Gten, and \Geleven\ 
the evaporation effect is strong and
most substructure is erased (see, for example, circles in Fig. 10, 
left panel). 
For \sig=0.5 and 0.1\cmgr\ (halos  \Gfour\ and \Gfive), the effect is 
less pronounced but the number of subhalos is still less than in the CDM 
MW-sized halos, although at the low velocity end ($\vmax = 20\ \kms$), 
the number of subhalos becomes larger. A similar 
behavior is seen in the halo \Gtwelve, where the effective \sig\ of 
the MW-sized halo is 
$\approx 0.2\cmgr$. The subhalo $\vmax-$function of the halo \Gfourteen\ 
lies slightly above of the CDM one. The $\vmax-$function of subhalos 
seems to be sensitive to the effective 
\sig\ of the host halo rather than that of the subhalos. In any case, halos 
\Gfour, \Gfive, \Gtwelve, and \Gfourteen\ have {\it at least} as 
many subhalos with 
$\vmax\ \simless 25\ \kms$ as the CDM MW-sized halo. Thus, we can conclude 
that the {\it substructure problem} (Klypin et al. 1999; Moore et al. 1999)
is not solved by a SIDM model with $\alpha\leq 1$ and reasonable values of 
\sign, at least at the low \vmax\ end of the velocity function.

The explanation of why the survival time of subhalos in a SIDM scenario 
is comparable or even longer than the one in the collisionless CDM 
cosmology resides probably in the fact that the host halo tidal force, 
responsible for the disruption of substructure along with dynamical 
friction, in the center is not as strong as in the CDM case. There 
may also intervene a numerical effect: the cores of subhalos, where 
the action of the self-interaction is important, are not resolved 
in the simulations in such a way that subhalo density profiles 
are very similar to those of CDM ones.
Notice, however, that this argument applies only to relatively small
effective \sig\ values: if \sig\ is very high, subhalo concentrations 
can be modified significantly by particle evaporation.

For completeness we have also plotted the cumulative $\vmax-$function for the
fiducial cluster-sized halo $Cl1$ with the different values of 
($\sign,\alpha$) (Fig. 10, right panel). The total number of subhalos 
within the SIDM cluster-sized halos (a sphere of $1.5\ \mpch$ radius was used)
is only slightly smaller than for the CDM halo, excepting the 
cases with ($\sign,\alpha$)= (3.0, 0.0) and (0.5, 0.0). It is interesting to notice 
that the evaporation constraint on the Fundamental Plane of elliptical galaxies 
within the clusters set namely a maximum value of $\sig \approx \ 0.5\ \cmgr$ 
(when $\alpha=0$) according to Gnedin \& Ostriker (2001).
Regarding the shape of the $\vmax-$function, in the range of 200-300 \kms, 
the SI cluster-sized halos have more subhalos than the corresponding CDM halo 
but the contrary happens at the low-\vmax\ end, SI cluster-sized halos have 
less subhalos than the CDM halo. This could be attributed to a longer survival 
time of massive subhalos. We do not expect a serious observational conflict 
related to the subhalo population in SIDM cluster-sized halos with 
($\sign,\alpha$)= (0.1,0.0), (0.5,1.0) and (1.0,1.0).

\section{Viability of the SIDM cosmogony}

Spergel \& Steinhard (2000) proposed the SIDM model in an attempt to 
overcome the apparent discrepancy between the cuspy halos predicted by 
CDM and the shallow halos inferred from observations of dwarf and LSB
galaxies (\cite{Moore94}; \cite{FP94}; \cite{Burkert95}; \cite{dBM97}). 
This discrepancy has been challenged by \cite{SMT2000}, \cite{vanDB00}, 
and van den Bosch \& Swaters (2001) who concluded that the spatial resolution 
of observed rotation curves was not sufficient to put constraints on 
cosmological models. However, more recent observational studies, using the 
highest sensitivity and spatial resolution in HI, H$\alpha$ and CO lines 
show that halos of dwarf and LSB galaxies are actually less concentrated 
in the center than the prediction of the CDM model (\cite{Cote00}; \cite{BAC01}; 
\cite{Bolato02}; \cite{AG02}; de Blok et al. 2001a, 2001b; \cite{Marchesini02}). 
Also, there are pieces of evidence of soft cores in normal disk galaxies 
(e.g., \cite{Corsini99}; \cite{BS01}; \cite{Salucci01}) and in elliptical 
galaxies (\cite{Keeton01}), although the evidence is less direct than in the 
case of dwarf and LSB galaxies.

Assuming the non-singular isothermal or pseudo-isothermal halo models,
several authors have tried to find scaling laws for halo cores. In a 
pionering paper, \cite{Kormendy90} derived the halo component parameters of 
some dSph and spiral galaxies, and concluded that halos with larger central 
velocity dispersions, $\sigma$, have larger core radii, r$_c$, and smaller 
central densities, $\rho_c$ (see also \cite{Burkert95}; \cite{SB00}). 
The correlations they found have a large scatter. On the other hand, for normal 
spirals, the halo parameters depend strongly on the assumed stellar M/L ratios, 
and for dSph galaxies the observational uncertainties are large.  
\begin{figure*}[ht]
\pspicture(0.5,-1.0)(15.0,11.2)
\rput[tl]{0}(1.0,10.7){\epsfxsize=18cm
\epsffile{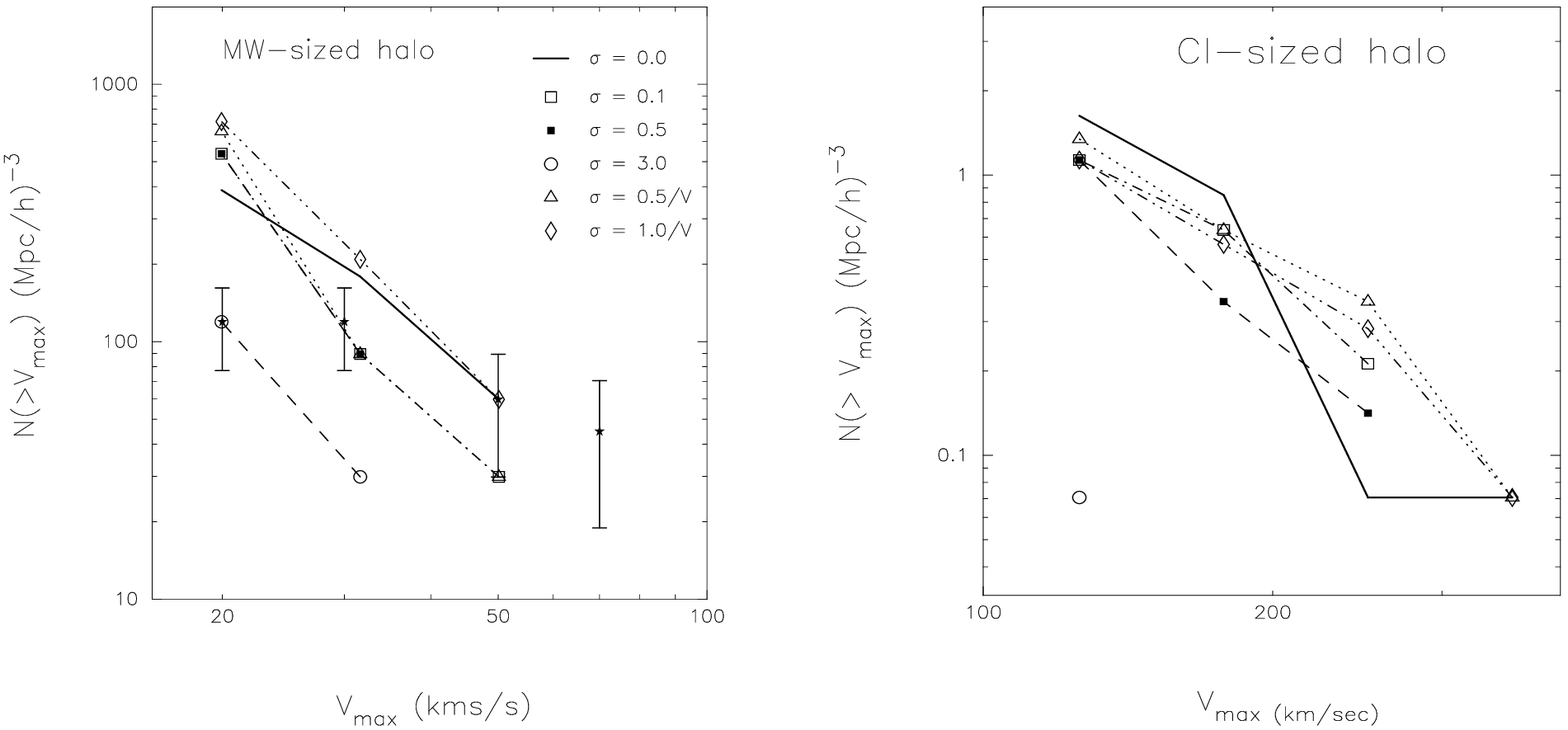}}
\rput[tl]{0}(0.5,1.5){
\begin{minipage}{18.4cm}
  \small\parindent=3.5mm {\sc Fig.}~10.---
Cumulative $\vmax-$function of MW- (left panel) and 
cluster-sized halos (right panel). All $G1$ ($Cl1$) subhalos with \vmax\ greater 
than 20 \kms\ (126 \kms) and within a sphere of radius 200 \kpch\ ($1.5 \mpch$) 
were counted. The averaged $\vmax-$function from satellites 
of MW and Andromeda is represented by stars (taken from \cite{KKVP99})
and it is shown in left panel. Error bars are just Poisson errors. The number
of subhalos in SI halos with a reasonable \sig\ (\ie, with a \sig\ that also
satisfies other observational constrains) is comparable to the one seen in 
the CDM prediction. Notice that there is only one subhalo within halo \clfive\ 
and in the last two \vmax\ bins of halo \clone. The presence of few more massive 
subhalos in SI cluster-sized
halos could be explained by a longer survival time of these halos with respect to
the survival time of subhalos in the CDM halo.
\end{minipage}}
\endpspicture
\end{figure*}

More recently, Firmani et al. (2000, 2001a), in an effort to infer the core
halo scaling laws, have analyzed a sample of dwarf and LSB galaxies with 
high-quality rotation curves and cluster of galaxies studied with gravitational 
lensing. They used the non-singular isothermal model and took into account 
the effect of gravitational drag by the baryon matter on the halo for the LSB 
galaxies. Firmani et al. found that: (i) $\rho_c$ exhibits a large scatter, 
while a correlation with the maximum circular velocity, \vmax\ 
(or $\sigma$), is not obvious, and (ii) r$_c$ and \vmax\ tend to correlate
with a slope smaller than 1. The different mass 
models of observational samples presented by several authors lead
to a similar conclusion (C\^ot\`e et al. 2000 for dIrr's; Verheijen 1997, 
and de Blok et al. 2001b for LSB galaxies).
All these authors use different M/L ratio assumptions (not too important for 
dwarfs) and apply either the non-singular isothermal or the pseudo-isothermal 
models. For the former case, $\rhosl=0.54\rho_c$ and $\rcsl=0.7$r$_c$, while for 
the latter $\rhosl=0.5\rho_c$ and $\rcsl=$r$_c$. We estimate $\rhosl$ and 
\rcsl\ from the Firmani et al. compilation and the works mentioned above. From 
Verheijen (1997) and de Blok et al. (2001b) we use their fits with the equal M/L 
ratio, and for the C\^ot\`e et al. (2000) data on nearby dwarfs we assume 
$h=0.65$. In the 
range of the observations,  $\vmax\sim 40-250$ kms/s, \rhosl\ 
oscillate roughly between $10^{16}$ and $10^{17}$ \densunit\ with the average at 
$\approx 3\times 10^{16} \densunit$. The radius \rcsl\ increases with \vmax\ 
but the scatter is very large; for \vmax=100 \kms, $\rcsl\approx 3 \kpch$
on average. Note that for the LSB galaxy data from Verheijen (1997) and
de Block et al. (2001b) the inferred central densities and core radii are upper and
lower limits, respectively. The formation of a central disk
contracts adiabatically the inner halo. For a typical LSB
galaxy, the halo core radius could have shrinked by a factor of $\sim 1.5$ 
after disk formation (Firmani \& Avila-Reese 2000; Firmani et al. 2001a),
increasing the central density by a factor of $\sim 2$.
For higher surface brightness galaxies these factors are larger. 
Observations show that the fitted halo central density indeed increases with 
disk surface brightness (Verheijen 1997; Swaters 1999). Taking into
account a correction for the further halo contraction, the scatter in the plot 
\rhosl-\vmax\ diminishes and the correlation between \rcsl\ and \vmax\ becomes
tighter. 

At the scale of cluster of galaxies, the highest resolution mass distribution 
analyses were done on the basis of either strong gravitational lensing studies 
or high-resolution X-ray studies of the intracluster medium.  For the cluster 
CL0024+1654 studied with strong gravitational lensing techniques (\cite{TKd98}), 
Firmani et al. (2001a) found a central density similar 
to those of their dwarf and LSB galaxy sample. From Firmani et al., we calculate
$\rhosl=2.6\times 10^{16} \densunit$ and $\rcsl=45.5\kpch$ (\vmax=2400\kms).
Most of the measured cluster mass distributions from high-resolution 
X-ray studies of cluster of galaxies (mainly with the Chandra satellite) are 
well fitted by both the NFW profile and the pseudo-isothermal 
model. From the fitting parameters reported for the clusters
A1835 (\cite{SAF01}), A2390 (\cite{AEF01}),
RXJ1347.5-1145 (\cite{ASF01}), and EMSS 1358+62 45 (\cite{ABG02}; 
this cluster is at $z=0.328$), we calculate
$\rhosl=2.8\times 10^{16}$, $2.0\times 10^{16}$, $9.1\times 10^{16}$ and
$\geq 2.4\times 10^{17} \ \densunit$, $\rcsl=32.5,\ 37.5, \ 22.5$ and
$< 6\ \kpch$, and $\vmax=1843,\ 1785,\ 2290$ and 1008\ \kms, respectively.  
Note, however, that the inclusion of the adiabatic cluster halo contraction 
due to a cD or another central galaxie would favor more extended shallow cores. 

The dotted lines in Figure 4 encompass roughly the 2$\sigma$ dispersion
of the observational inferences of \rhosl\ and \rcsl\ versus \vmax\ for
the dwarf and LSB galaxies and the cluster of galaxies described above. The
dashed lines are lines representative of the average behavior of the data. 
In Figure 4 were also plotted the model predictions for different values
of $\sig$. At cluster scales the $\Lambda$CDM halos seem to 
be in agreement with observations. However, at galaxy scales they are
roughly a factor of ten denser at the radius \rcsl. Extrapolating to
scales of dwarf galaxies this factor would be even larger.
On the other hand, a SIDM model with an effective $\sig$ as large as 
3.0 \cmgr produces halos with a large variety of inner structures. As 
discussed in \S 3, for these values of $\sig$ the core collapse 
is triggered in a time scale smaller than the Hubble time.
However, we also have seen that 
the core collapse could  be delayed by late major mergers or even stopped 
by a continuos injection of thermal energy from the hotter particles of 
the ambient in which the halo is embedded. Thus, in a cosmogony with a
relatively high effective cross section, some halos could show
shallow inner density profiles (e.g., elliptical and 
dwarf galaxy halos), while those which are isolated would be in the
core collapse phase showing cuspy inner density profiles
(e.g., LSB galaxy halos). Nevertheless, the major conflict in this case is that
substructure is strongly erased; for example, for $\sig=3 \cmgr$, most
subhalos in cluster- and galaxy-sized halos are evaporated,
in complete disagreement with observations (Fig. 10).

The SIDM simulations with  $(\sign,\alpha) = (0.1,0.0)$ and $(0.5,0.0)$ show that 
at galaxy scales the central densities of the halos agree roughly with observations, 
the latter case producing probably slightly shallower halos than observed. However, 
at the cluster scales the halos in both cases are too shallow w.r.t. 
observations. We are then left with SIDM models in which \sig\ varies as the 
inverse of \vrel. For the cases explored here, $(\sign,\alpha) = (0.5,1.0)$ and 
(1.0,1.0), halos at galaxy and cluster scales fall within the obsevational range, 
and $\rho_c$ {\it and the halo soft core fraction are approximately constant 
with scale (\vmax).} Unfortunately, we can not explore the inner density 
profiles of the (sub) halos at the scales of dwarf galaxies because of they
are resolved with too few particles.
Extrapolating, it seems that halos at the scales of dwarf galaxies 
would remain as shallow as the larger ones.
A weak hint that this could be happening comes from meassuring the 
c$_M$ concentration parameter for subhalos of MW-sized halos. This
parameter characterizes the shape of the density profiles at radii 
$\sim 0.03-0.1$R$_{vir}$. In the central panels of Figure 7 one sees that 
c$_M$ for subhalos of halos \Gtwelve, \Gthirteen, and \Gfourteen\ 
is on average lower than that corresponding to CDM subhalos. 
To a first approximation, we expect $\rho_c$ to be independent of \vmax\
when $\sig\propto 1/\vrel$ because \ncol\ (eq. 4) or the collisional 
probability $P$ do not depend in this case on $\vrel$. 
However, \ncol\ or $P$ also depend on the local density and this is typically 
larger for smaller halos. 

The satellite cumulative $\vmax-$function of the halo $G1$ with
$(\sign,\alpha) = (0.5,1.0)$ is slightly higher than the one of 
the corresponding CDM halo at the low velocity limit, and then drops by
a big factor, giving good agreement with observations for $\vmax\gtrsim 30\kms$ 
(Fig. 10). For the $(\sign,\alpha) = (1.0,1.0)$ case, the function lies
above of the CDM prediction up to \vmax=50 \kms. As explained
in \S 4, this could be due to the softer inner gravitational potential of the
SIDM halos with respect to the CDM ones. For the explored range of the SI
parameter space, it seems that the main influence on the number and structure of
subhalos comes from the scattering properties of the host halo rather than
those intrinsic to subhalos (collisions only between subhalo particles). In \S 3.2
we have seen indeed how structures of subhalos with similar
\sig\ differ because the effective \sig\ of the host halos are different.    

In conclusion, among the SIDM models studied in this paper, those with \sig\ 
inversely proportional to \vrel\ are able to produce halo cores shallower than
CDM and with central densities roughly constant with \vmax. However, the 
substructure problem for MW-sized halos is not solved. 
Reionization has been proposed as a mechanism able to inhibit
the formation of dwarf galaxies within the large population of predicted
small CDM halos (Bullock et al. 2000), in such a way that the substructure
excess would not be longer a critical issue. Nevertheless, a potential problem
with the Tully-Fisher relation of dwarf galaxies would remain (Avila-Reese
et al. 2001).

Recently, HO carefully analized several observational
constraints for the SIDM model. Their study is mostly based on analytical
estimates calibrated on basis of previous numerical results. These estimates
agree with the results obtained here for SIDM models with a constant \sig\ and 
with a \sig\ inversely proportional to \vrel. If we add the constrain given
by the satellite overabundance in MW-sized
halos, then the region of allowed $(\sign,\alpha)$ values  
becomes smaller than in HO since models with $\alpha\lesssim
1$ also would be ruled out. However, as mentioned above, this constrain is 
weaker than the others because the satellite excess problem could be solved 
invoking reionization.

HO have introduced one extra constrain, namely the formation 
of supermassive black holes (SMBH) due to SI (Ostriker 2000). 
According to their estimates and using the observed mass of the central 
SMBH of our Galaxy and the lack of a SMBH in the bulge-less galaxy M33, they
conclude that $\sig$ should be smaller than $0.02\ \cmgr$ for any value of
$\alpha\leq 1$. A critical assumption of this constrain is that
the dark halo initially has the typical cuspy CDM density profile, i.e., the 
establishment of the soft core by SI occurs only at time scales 
close to a Hubble time. In fact,
the mass of the SMBH formed by accretion of SIDM is extremely sensitive 
to the inner halo density profile slope. For a slope shallower than $-1$
this mass is much less than $10^3\msun$ for $\alpha\leq1$ (see Fig. 1 in 
HO). Our simulations show that the evolving time scales 
under SI for the cases analyzed here are much less
than a Hubble time. The core expansion phase actually occurs in time scales 
close to the halo dynamical time (Figs. 1 and 5) for $\sig\sim 1\ \cmgr$. The 
calculations of halo evolution under SI of Firmani et al. (2001b) also show 
that the inner halo density profile flattens quickly, as the halo grows and 
virializes. So, accretion of SIDM could not be the relevant mechanism for
the formation of SMBH and, therefore, the strong constraint of HO
might not be valid. For the formation of SMBH other alternative mechanims
have been proposed (e.g., Granato et al. 2001).

\section{Conclusions}

A set of high-resolution N-body simulations has been run to study 
the structure and substructure of cluster- and MW-sized halos in a 
SI $\Lambda$CDM cosmology with cross sections $\sig = \sign (1/\vcien)^\alpha$, 
where \sign\ is in units of \cmgr, and \vcien\ is the relative velocity 
in units of 100 \kms. Our main conclusions follows:

% Core collapse
-For low values of \sig, when $\ncol\ \lesssim 2-5$ at $z = 0$, and for a
given mass, the lower the \sig, the smaller the core radii and 
the higher the central densities. This behavior may be reversed for high 
values of \sig. In this last case, the evolution of halos under SI is fast 
and the core collapse phase could begin well before $z = 0$. We analysed 2 
cluster- and 3 MW-sized halos with $(\sign,\alpha) = (3.0, 0.0)$ and different 
mass assembly histories. We have found that they go through the 
core collapse phase, although this phase can be delayed significantly for 
some halos by both dynamical heating due to {\it major mergers} (but not smooth 
mass accretion) and the evaporation of (sub)halo particles interacting with 
the hot particles of the host halo.

%the physics of SIDM

-The SIDM halos expand their cores due to a heat inflow from the 
hotter surroundings. Since the core heat capacity, $C$, is positive, this 
process leads to the isothermalization of the core.  After the core maximum 
expansion, $C$ becomes negative down to smaller radii. Then the gravothermal 
instability triggers and the core collapses. The system moves from a minimum to 
a maximum entropy state. If the halo periphery is heated and the temperature 
becomes higher than in the core, the heat inflows to the core and the core expands
until the overall halo isothermalizes. This process delays the core collapse, 
but will hardly reverse the gravothermal instability to a runaway core expansion.

% Central density (isolated halos)

-Present day halos with $\ncol\ \lesssim\ 2-5$ are still expanding their cores
or have just had the gravothermal catastrophe triggered. The inner density 
profiles when isolated (host) are flattened by SI only in the innermost 
regions, $r < 0.05\rvir$. When \sig\ is constant, massive halos are suposed 
to be the most affected by SI ($\ncol \propto \rho \vmax$). For typical 
cluster-sized halos to have $\ncol\ \lesssim\ 2-5$, \sig\ should be smaller 
than $\approx 0.5\ \cmgr$. For $\sig\ \lesssim 0.5\ \cmgr$, the lower the 
\vmax, the higher the $\rho_c$. On the other hand, when $\sig \propto 1/\vrel$, 
this trend is much weaker: $\rho_c$ remains almost constant with \vmax. We
compare the predicted core densities and radii as well as the scaling laws of 
the halo cores with those inferred from the observations of dwarf and LSB galaxies 
and cluster of galaxies. The SIDM models with $\sig\ = 0.5-1.0\ (1/\vcien)\ \cmgr$ 
are favored. With these values of \sig, \ncol\ is smaller than 2-5 from 
galaxy- to cluster-sized halos.

% Structure and population of subhalos
 
-We find that the structure and population of subhalos are determined 
by the interaction between the hot host halo particles and the cooler 
subhalo particles rather than by internal processes in the subhalos. Our 
simulations show that overall SIDM subhalos become puffier than their CDM 
counterparts and that they are completely evaporated in the limit of high 
\sig. For all the ($\sign,\alpha$) values used here, we do not see signs 
of a core collapse in subhalos, but we warn that this result can be masked 
by the poor mass resolution of the simulations at the subhalos scales. 

-The number of subhalos within MW-sized halos is largely suppresed for 
($\sign,\alpha$) = (3.0,0.0). The cumulative $\vmax-$functions of models
($\sign,\alpha$) = (0.1,0.0),  (0.5,0.0) and (0.5,1.0) agree roughly with 
the observations for $\vmax \gtrsim 30 \kms$, but at the low \vmax\ end 
they lie slightly above the corresponding CDM one. For ($\sign,\alpha$) = 
(1.0,1.0), this function is higher than the CDM at all \vmax. Subhalos may 
survive a longer time in SIDM halos than in CDM ones because the inner 
tidal force in the former is weaker than in the latter. Our results show 
that the range of ($\sign,\alpha$) values in which the SIDM
scenario is interesting does not solve the potential problem of
the excess of substructure in MW-sized halos.
 
% Ellipticity y momento angular
 
- The SIDM halos have more angular momentum at a given mass shell
than their CDM counterparts. However, this extra angular momentum is small 
and does not solve the conflict between the measured angular momentum 
distribution in CDM halos and that infered from observations
 for dwarf galaxies. On the other hand, the SIDM core halos are indeed rounder 
than their CDM counterparts, but they are not spherical. This roundness is not 
only seen in the core, it can extend all the way to the virial radius of the halo.

We conclude that the problem of cuspy halos in the $\Lambda$CDM cosmology may 
be solved by introducing a modest dark matter cross section inversely proportional 
to the relative velocity of interaction, $\sig\approx 0.5-1.0 (1/\vcien)\ \cmgr$. 
On the other hand, our simulations show that for these values of \sig, the number 
of subhalos in MW-sized halos is close to that of the CDM counterparts. Our
simulations also show that SI flattens the inner density profiles of growing CDM 
halos since early epochs in such a way that the grow of super massive black holes
by accretion of SIDM is not an efficient process. A remaining interesting question 
is whether the CDM candidate particles may be predicted with the cross sections
after their decoupling prefered by our results.

\acknowledgments
 
We are grateful to Anatoly  Klypin and Andrey Kravtsov for kindly 
providing us a copy of the ART code in its version of multiple mass, and for
our enlightening discussions. The authors thank also the anonymous referee 
for constructive suggestions. P.C. would like to thank A. Watson and 
W. Henney for the use of a PC in which part of the ART simulations 
were done. Part of the ART simulations were also performed at the 
Direcci\'on General de Servicios de C\'omputo Acad\'emico, UNAM, using 
an Origin-2000 computer. This work was supported by CONACyT grants
33776-E to V.A. and 36584-E to P.C.

%===========================

\end{document}